\documentclass[preprint,prb,aps]{revtex4}
\usepackage{amsmath,amssymb}
\usepackage{graphicx}% Include figure files
\usepackage{dcolumn}% Align table columns on decimal point
\usepackage{bm}% bold math

\begin{document}
\newcommand{\ds}{\displaystyle}
\newcommand{\bea}{\begin{eqnarray}}
\newcommand{\eea}{\end{eqnarray}}

%\preprint{APS/123-QED}

\title{A Group Theoretical Identification of Integrable Equations in
the Li\'enard Type Equation $\ddot{x}+f(x)\dot{x}+g(x) = 0$ : Part II:
Equations having Maximal Lie Point Symmetries}
% Force line breaks with \\
\author{S. N. Pandey}%
\email{snp@mnnit.ac.in (S. N. Pandey)}
\affiliation{%
Department of Physics, Motilal Nehru National Institute of
Technology, Allahabad - 211 004, India}%
\author{P. S. Bindu, M. Senthilvelan and M. Lakshmanan}%
\email{lakshman@cnld.bdu.ac.in (M. Lakshmanan)}
\affiliation{%
Centre for Nonlinear Dynamics, School of
Physics, Bharathidasan University,
 Tiruchirapalli -620 024, India
}%
\date{\today}
\begin{abstract}
In this second of the set of two papers on Lie symmetry
analysis of a class of Li\'enard type equation of the form $\ddot {x}
+ f(x)\dot {x} + g(x)= 0$, where over dot denotes differentiation
with respect to time and $f(x)$ and $g(x)$ are smooth functions
of their variables,  we isolate the equations which
possess maximal Lie point symmetries.  It is well known that
any second order nonlinear ordinary differential equation which
admits eight parameter Lie point symmetries is linearizable to free
particle equation through point transformation.  As a consequence
all the identified equations turn out to be linearizable.  We also show that
one can get
maximal Lie point symmetries for the above Li\'enard equation only when
$f_{xx} =0$ (subscript denotes differentiation).  In addition,
we discuss the linearising transformations and
solutions for all the nonlinear equations identified in this
paper.
\end{abstract}

\maketitle

\section{\bf Introduction}
\label{sec1}

The present paper continues the
work on the classification of Lie point symmetries described in the previous
paper\cite{pandey:2007} of the second order ordinary differential equation
of the Li\'enard type,
\begin{eqnarray}
A(x,\dot{x},\ddot{x})\equiv \ddot {x} + f(x)\dot {x} + g(x)= 0,
\label{1.1}
\end{eqnarray}
where $f$ and $g$ are smooth functions of $x$ and over-dot
denotes differentiation with respect to $t$.  In the first part\cite{pandey:2007}
while solving the determining equations we have assumed the case in
which one of the symmetry functions $b(x) = 0$ and identified the corresponding
equations.  The question now naturally
arises as to what are the invariant equations when $b(x) \neq 0$. In
this part we present the answer to this question.

On solving the determining equations with $b(x) \neq 0$ we find that
a class of equations possess eight parameter Lie-point
symmetries.  The forms of $f$ and $g$ which lead to the maximum
number of symmetry generators are as follows (in increasing order of
generality):
\begin{align}
\mbox{(i)~~} & f  =  0, &&  g  =  \lambda_1x+\lambda_2,
\nonumber\\
\mbox{(ii)~~} &  f  =  k, &&  g  = \lambda_1x+\lambda_2,\nonumber\\
\mbox{(iii)~~} &  f  =  kx, &&   g  = \frac{1}{9}k^2 x^3+\lambda_1x +
\lambda_2,  \nonumber\\
\mbox{(iv)~~} & f  =  k_1x+k_2, &&  g  =  \frac{1}{9}k_1^2x^3
+\frac{1}{3}k_1k_2x^2+\lambda_1x +\lambda_2,
\label{s1}
\end{align}
where $k,k_1,k_2,\lambda_1$ and $\lambda_2$ are constants.  The
first case includes the equation of motion of the
free particle and simple
harmonic oscillator. It is well known that both the equations
admit eight symmetry generators \cite{agu:1984,Hydon:2000,wulf:1976} satisfying an
$sl(3,R)$ algebra.
Similarly, the case (ii) which corresponds to the damped harmonic
oscillator which has also been shown to admit $sl(3,R)$ symmetry algebra
\cite{cervil:1984a}. From case (iii), by restricting the
parameters $\lambda_1=\lambda_2=0$, one can get the modified Emden
type equation which is the first nonlinear ODE shown by Mahomed and Leach
to admit a rich set of Lie point symmetry generators\cite{mahomed:1985}.
Here, we show that the general modified Emden type equation
with linear term and constant external forcing also admits eight
point symmetry generators. We also report the class of equations
which falls under case (iv) admitting eight symmetry generators
and this may be considered as the most general form of
(\ref{1.1}).  It is important to note that  all these equations
are linearizable under point transformations.  The explicit forms of the infinitesimal
symmetries and their associated generators for the general cases
are also reported for the first time, as far as the authors'
knowledge goes.

In all the above
cases,  it turns out that $f_{xx} =0$.  By considering the next higher
degree polynomial in the variable $x$ for $f$, say quadratic, $
f  = k_1x^2+k_2x+k_3$, so that $f_{xx} \neq 0$, we get
\begin{eqnarray}
g  = \frac{1}{9}k_1^2 x^5+\frac{5}{18}k_1
k_2x^4+\lambda_1x^3 +\lambda_2x^2+\lambda_3x+\lambda_4,\nonumber
\end{eqnarray}
which leads to one symmetry generator only.  In general, we
prove that for $f_{xx} \neq 0$
one obtains lesser parameter symmetry group only.

The plan of the paper is as follows. In Sec.~II we recall the
derivation of the determining equations for the infinitesimal
symmetries and describe a procedure to solve them.  In Sec.~III by
demanding all the four symmetry functions to be non-zero, we construct
systematically the class of functions $f$ and $g$ which admit
eight symmetry generators.  We start our analysis with
$f=0, g=0$ in Eq. (\ref{1.1}), which lead to the equation of motion of the free
particle, and derive all earlier known equations from (\ref{1.1})
which admit eight point symmetries. Proceeding further by
assuming the function $f$ to be  a constant we deduce damped
harmonic oscillator equation and its variant equations.  We also
present the infinitesimal symmetries and their associated generators
for these equations.  Next, in Sec. IV, by considering the function $f$ to be
linear in $x$, we deduce two new
family of nonlinear ODEs which admit eight symmetry generators.  In
Sec. V, we prove from the symmetry determining equations that when $f_{xx} \neq 0$
only lesser parameter symmetries
can exist.  We also show the equivalence of our results
with the Lie's linearization criterion.
Finally in Sec.~VI we provide a summary of our results. The solutions
of the determining equations encountered in the symmetry analysis of
new equations are presented in Appendices A and B.
\section{\bf Symmetry determining equation of (\ref{1.1})
with $ {b \neq 0}$} \label{ec3}

To make this paper self contained we recall the essential determining
equations very briefly here.  The invariance of Eq. (\ref{1.1}) under
one parameter Lie group of infinitesimal transformations,
$\tilde{t}=t+\epsilon \xi(t,x)+O(\epsilon^2), \;\;
\tilde{x}=x+\epsilon \eta(t,x)+O(\epsilon^2), \;\epsilon \ll 1,$
leads us to four determining equations for the functions $\xi$ and $\eta$
(vide Eqs. (8)-(11) in Paper I).  The
first two determining equations can be integrated straightforwardly to yield
$\xi = a(t)+b(t)x, \;\;
\eta = \dot{b}{x}^2-2b\Im(x) +c(t)x+d(t)$,
where
$\Im_x = F(x) = \int_0^{x}f(x')dx'\;\mbox{and}\;
\Im_{xx}=f(x)$.  The
remaining two determining equations, after substituting the forms of
$\xi$ and $\eta$,  read
\begin{eqnarray}
(\dot{b} x^2-2b\Im+cx+d)f_x+(\dot{a}+\dot{b}x)f+3bg-4\dot{b}F+
3\ddot{b}x+2\dot{c}-\ddot{a}=0, \label{2.14}
\end{eqnarray}
and
\begin{eqnarray}
&&(\dot{b}x^2-2b\Im+cx+d)g_x-(c-2\dot{a}-2bF)g - 2\ddot{b}\Im
\nonumber\\
%\hspace*{0.1in}
&&\qquad\qquad\qquad\qquad
+(\ddot{b} x^2-2\dot{b}\Im + \dot{c}x
+\dot{d})f +\dddot{b}x^2+\ddot{c}x +\ddot{d}  = 0. \label{2.15}
\end{eqnarray}
Since $g$ is linear in (\ref{2.14}) we can rewrite (\ref{2.14})
to obtain
\begin{eqnarray}
g=\frac{1}{3b} [-(\dot{b}x^2-2b\Im+cx+d)f_x-(\dot{a}+\dot{b}x)f
+4\dot{b}F-3\ddot{b}x-2\dot{c}+\ddot{a} ], \; b \neq 0.
\label{2.16}
\end{eqnarray}
While deriving Eq. (\ref{2.16}) we assumed that $b\neq 0$.  On the
other hand one may also consider the case $b=0$. In fact this was
the case of the Paper I.  Now we solve the determining equations
(\ref{2.14}) and (\ref{2.15}) with $b\neq 0$ and identify the functions
that are invariant under eight parameter Lie point symmetries.
\section{\bf Maximal Lie point symmetries of Li\'enard  type systems: Linear
ODES }
To begin with we consider polynomial forms for $f$ in the variable $x$ with
increasing degrees and derive the
corresponding explicit forms of $g$ through Eq. (\ref{2.16}). Substituting
the expressions
of $f$ and $g$ into Eq. (\ref{2.15}) and
solving the resultant equation we obtain the infinitesimal
symmetries $\xi$ and $\eta$
which in turn leads to the classification of the class of second order ODEs
which admit maximum number of symmetry generators.  We find that the
maximal number (eight) of symmetries exists only when  $f_{xx}=0$.
When $f_{xx} \neq 0$, lesser parameter Lie point symmetries only exist.
\subsection{\bf Linear Undamped Systems: $ f=0$}
To start with we consider the simple form $f=0$. Correspondingly from Eq.
(\ref{2.16}) we get
\begin{eqnarray}
g=\lambda_1 x +\lambda_2,
\label{3.1}
\end{eqnarray}
where $\lambda_1$ and $\lambda_2$ are arbitrary constants.
Substituting Eq.~(\ref{3.1}) into (\ref{2.15}) and solving it consistently
we can find the infinitesimal
symmetries. However, to classify the results systematically we consider the
following subcases, \emph{viz.}, $(i)~\lambda_1$ = $\lambda_2$ = $0$,
$(ii)~\lambda_2 \neq 0$,~$\lambda_1$ = $0$,
 $(iii)~\lambda_1 \neq 0,~ \lambda_2$ = $0$ and
$(iv)~\lambda_1 \neq \lambda_2 \neq 0$.
\subsubsection{\bf Free particle motion ($\lambda_1 = \lambda_2 = 0)$}
When both $\lambda_1$ and $\lambda_2$ are equal to zero, the
function $g$  also becomes zero and so Eq. (\ref{1.1})
takes the form
\begin{eqnarray}
\ddot{x}=0
\label{3.2}
\end{eqnarray}
which is obviously the equation of motion of a free particle. Substituting $f
= g = 0$ into Eq. (\ref{2.15}) we obtain the equation $\dddot{b}x^2+
\ddot{c}x+\ddot{d}=0$.  Equating the coefficients of $x^i,\;i=0,1,2$, to zero
separately one
obtains $\dddot{b}=0$, $\ddot{c}=0$ and $\ddot{d}=0$.  Solving the later
two equations
one obtains $c=c_1+c_2t$ and $d=d_1+d_2t$, where $c_i's$ and
$d_i's,\; i=1,2$, are arbitrary constants.  Similarly from Eq. (\ref{2.14}),
we obtain $\ddot b=0$ and $\ddot a=2\dot c$. Solving these two equations
one obtains $b=b_1+b_2t$ and $a=a_1+a_2t+c_2t^2$, where $b_i's$ and
$a_i's,\; i=1,2$, are arbitrary constants. Inserting the forms
$a(t),b(t),c(t)$ and $d(t)$ in $\xi = a(t)+b(t)x$
and $\eta=\dot{b}{x}^2+c(t)x+d(t)$ one gets
\begin{eqnarray}
\xi =  a_1+a_2t+c_2t^2+(b_1+b_2t)x,\;\;
\eta  =  d_1+d_2t+(c_1+c_2t)x+b_2x^2.
\label{3.3}
\end{eqnarray}
The associated eight infinitesimal generators take the following form
\begin{eqnarray}
X_1 = \frac{\partial}{\partial t}, \;\;
X_2 = t\frac{\partial}{\partial t},\;\;
X_3=x\frac{\partial}{\partial t},\;\;
X_4 = x\left(t\frac{\partial}{\partial t}
+x\frac{\partial}{\partial x}\right),
\nonumber\\
X_5 = x\frac{\partial}{\partial x},\;\;
X_6 = t\left(t\frac{\partial}{\partial t}+x\frac{\partial}{\partial
x}\right),\;\;
X_7 = \frac{\partial}{\partial x},\;\;
X_8 = t\frac{\partial}{\partial x},
\label{3.4}
\end{eqnarray}
which satisfy an $sl(3, R)$ algebra.  The symmetries and their
generators (\ref{3.4}) coincide exactly with the known ones
given in Refs.\onlinecite{agu:1984,Hydon:2000}.
\subsubsection{\bf Free falling particle ($\lambda_2,\;\neq 0,\; \lambda_1 = 0)$}
If we choose $\lambda_1=0$ and $\lambda_2$ as arbitrary in Eq. (\ref{3.1}), then from
Eq. (\ref{1.1}) we get
\begin{eqnarray}
\ddot{x}+\lambda_2=0,
\label{3.6}
\end{eqnarray}
which is the equation of motion of a free falling particle. Now solving the
determining Eq.
(\ref{2.15}) with $f = 0$ and $g = \lambda_2$ we obtain \cite{agu:1984}
\begin{eqnarray}
\xi  & = & a_1+a_2t+(c_2+\frac{3}{2}\lambda_2b_1)t^2+
\frac{\lambda_2}{2}b_2t^3+(b_1+b_2t)x, \nonumber\\
\eta & = & d_1+d_2t+\frac{\lambda_2}{2}(c_1-2a_2)t^2-
\lambda_2(\lambda_2b_1+\frac{1}{2}c_2)t^3
-\frac{\lambda_2^2}{4}b_2t^4
+(c_1+c_2t)x+b_2x^2,
\label{3.7}
\end{eqnarray}
where $a_i,b_i,c_i$ and $d_i,\; i=1,2$, are arbitrary constants.
Here too, we have the following eight generators \cite{agu:1984}
\begin{eqnarray}
\quad
X_1 &  =&   \frac{\partial}{\partial t},\;\;
X_2 = t\frac{\partial}{\partial t}-\lambda_2t^2\frac{\partial}
{\partial x}, \;\;
X_3 = (x+\frac{3}{2}\lambda_2t^2)\frac{\partial}{\partial t}
-\lambda_2^2t^3\frac{\partial}{\partial x},
\nonumber\\\quad
X_4  & = &  t(x+\frac{\lambda_2}{2}t^2)\frac{\partial}{\partial t}
+(x^2-\frac{\lambda_2^2}{4}t^4)\frac{\partial}{\partial x},\;\;
X_5 = (x+\frac{\lambda_2}{2}t^2)\frac{\partial}{\partial x},
\nonumber\\
 \quad
X_6  & = &  t^2\frac{\partial}{\partial t}+
t(x-\frac{\lambda_2}{2}t^2)\frac{\partial}{\partial x},\;\;
X_7=\frac{\partial}{\partial x},\;\;
X_8=t\frac{\partial}{\partial x},
\label{3.8}
\end{eqnarray}
which also form an $sl(3, R)$ algebra.
\subsubsection{\bf Free linear harmonic oscillator ($\lambda_1
\neq 0,~\lambda_2 = 0)$}
We consider $\lambda_1 $ as arbitrary and $\lambda_2 = 0$
so that Eq. (\ref{1.1}) becomes
\begin{equation}
\ddot{x}+\lambda_1x = 0,
\label{3.10}
\end{equation}
which is nothing but the equation of a linear harmonic oscillator (for
$\lambda_1 >0$).  Solving Eq. (\ref{2.15}) with $f=0$ and $g=\lambda_1x$ we
get the following well known infinitesimal symmetry
transformations \cite{agu:1984}
\begin{eqnarray}
\xi&=&a_1+a_2\sin 2\alpha t+a_3\cos 2\alpha t
+(b_1\sin \alpha t+b_2\cos \alpha t)x,  \label{3.11a}\\
\eta&=&d_1\sin \alpha t+d_2\cos \alpha t+(c_1+a_2\alpha\cos 2\alpha t
-a_3\alpha\sin 2\alpha t)x
+\alpha(b_1\cos\alpha t-b_2\sin \alpha t)x^2, \nonumber
\end{eqnarray}
for Eq. (\ref{3.10})
where $\alpha^2 = \lambda_1$ and
$a_i,b_j,d_j$,$\;i = 1,2,3$, $j = 1,2$ and $c_1$ are arbitrary constants.
The associated generators forming the $sl(3,R)$ algebra are \cite{agu:1984}
\begin{eqnarray}
 X_1 &  = &  \frac{\partial}{\partial t}, \;\;
X_2 = \sin 2\alpha t\frac{\partial}{\partial t}+\alpha x\cos 2\alpha t
\frac{\partial}{\partial x},
 X_3  =   \cos 2\alpha t\frac{\partial}{\partial t}-\alpha x\sin 2\alpha t
\frac{\partial}{\partial x},\nonumber \\
X_4 & = & x(\sin \alpha t\frac{\partial}{\partial t}
+\alpha x\cos \alpha t)
\frac{\partial}{\partial x},
 X_5  =  x(\cos \alpha t\frac{\partial}{\partial t}
 -\alpha x\sin \alpha t)
\frac{\partial}{\partial x},\nonumber \\
X_6 & = & x\frac{\partial}{\partial x},\;\;
 X_7  =  \sin \alpha t\frac{\partial}{\partial x},\;\;
X_8 = \cos \alpha t\frac{\partial}{\partial x}.
\label{3.12}
\end{eqnarray}
On the other hand in the case $\lambda_1<0$, Eq. (\ref{3.10}) becomes the
repulsive harmonic oscillator which is invariant under the following
infinitesimal symmetries
\begin{eqnarray}
 \quad
\xi & =  & a_1+a_2e^{2\alpha t}+a_3e^{-2\alpha t}
+(b_1e^{\alpha t}+b_2e^{-\alpha t})x, \nonumber \\
 \quad
\eta  & = &  d_1e^{\alpha t}+d_2e^{-\alpha t}+(c_1+\alpha(a_2e^{2\alpha t}
-a_3e^{-2\alpha t}))x
+\alpha(b_1e^{\alpha t}-b_2e^{-\alpha t})x^2,
\label{3.11}
\end{eqnarray}
The associated $sl(3,R)$ symmetry generators are
\begin{eqnarray}
 X_1 &  = &  \frac{\partial}{\partial t}, \;\;
X_2 = e^{2\alpha t}(\frac{\partial}{\partial t}+\alpha x
\frac{\partial}{\partial x}),\;\;
 X_3  =   e^{-2\alpha t}(\frac{\partial}{\partial t}-\alpha x
\frac{\partial}{\partial x}),\;\;
X_4  =  xe^{\alpha t}(\frac{\partial}{\partial t}+\alpha x
\frac{\partial}{\partial x}),\nonumber \\
 X_5 & = & xe^{-\alpha t}(\frac{\partial}{\partial t}-\alpha x
\frac{\partial}{\partial x}),\;\;
X_6   =  e^{\alpha t}\frac{\partial}{\partial x},\;\;
 X_7  =  e^{-\alpha t}\frac{\partial}{\partial x},\;\;
X_8 = x\frac{\partial}{\partial x}.
\label{3.12}
\end{eqnarray}

\subsubsection{\bf Displaced linear harmonic oscillator
($\lambda_1\neq \lambda_2 \neq 0)$}
Finally, we consider  both $\lambda_1 $ and $\lambda_2$ to be
arbitrary (not equal to zero) and  we have an equation of the
form
\begin{equation}
\ddot{x}+\lambda_1 x +\lambda_2 = 0,
\label{3.14}
\end{equation}
which corresponds to a displaced simple harmonic motion.  Eq.~(\ref{3.14})
can be transformed to (\ref{3.10}) by a simple transformation,
$x \rightarrow x' $$= x+(\frac{\lambda_2}{\lambda_1})$ and consequently the
infinitesimal symmetries of the former can be derived from that of Eq.~(\ref{3.11a}).

\subsection{\bf Linear Damped Systems: $f = constant = k$}
In the previous subsection, III A, we considered the case $f=0$ and derived the form of $g$.
Now we fix $f=\mbox{constant}=k$ and deduce the associated form of $g$.
Substituting  $f = k$ in Eq. (\ref{2.16}), we again obtain $g$ as a linear function
in $x$, that is
\begin{eqnarray}
g=\lambda_1x+\lambda_2,
\label{3.15}
\end{eqnarray}
where $\lambda_1$ and $\lambda_2$ are arbitrary constants. Consequently one
can solve Eq. (\ref{2.15}) with (\ref{3.15}) to obtain the infinitesimal symmetries. However,
to relate the present results with those of
the literature, we consider the following four cases separately:
$(i)~\lambda_1 = \lambda_2 = 0$,
$(ii)~\lambda_1 \neq 0,\;\lambda_2 = 0$, $(iii)~\lambda_2 \neq
0,\;\lambda_1 = 0 , $ and $(iv)~\lambda_1$$\neq \lambda_2 \neq 0$.

\subsubsection{\bf Free particle in a viscous medium ($\lambda_1 =
\lambda_2 = 0)$}
Let $\lambda_1=\lambda_2=0$. Then from Eq. (\ref{3.15})  we obtain  $g = 0$.
Substituting $f = k$ and $g = 0$ in Eq. (\ref{1.1}), we get
\begin{equation}
\ddot{x}+k\dot{x}=0,
\label{3.16}
\end{equation}
which corresponds to the equation of a free particle in a viscous medium. Solving
Eq. (\ref{2.15}) with $f=k$ and $g=0$ we get
\begin{eqnarray}
\xi & = & a_1+\frac{a_2}{k}e^{kt}+\frac{c_2}{k^2}e^{-kt}
+(b_1+\frac{b_2}{k}e^{kt})x, \nonumber\\
\eta & = & d_1+\frac{d_2}{k}e^{-kt}+(c_1-\frac{c_2}{k}e^{-kt})x
-kb_1x^2,
\label{3.17}
\end{eqnarray}
where $a_i$, $b_i$, $c_i$ and $d_i$, $i=1,2$, are arbitrary constants.
The corresponding generators satisfying the  $sl(3,R)$ algebra are\cite{prince:1979}
\begin{eqnarray}
 X_1 & = & \frac{\partial}{\partial t},\;\;
 X_2 = \frac{1}{k}e^{kt}\frac{\partial}{\partial t}, \;\;
 X_3  =   x\left(\frac{\partial}{\partial t}
-kx\frac{\partial}{\partial x}\right),\;\;
 X_4 =  \frac{x}{k}e^{kt}\frac{\partial}{\partial t},\nonumber\\
 X_5 & = & x\frac{\partial}{\partial x},\;\;
 X_6 =  \frac{1}{k}e^{-kt}\left(\frac{1}{k}\frac{\partial}{\partial t}
-x\frac{\partial}{\partial x}\right),\;\;
 X_7 =  \frac{\partial}{\partial x}, \;\;
 X_8  =  \frac{1}{k}e^{-kt}\frac{\partial}{\partial x}.
\label{3.18}
\end{eqnarray}

\subsubsection{\bf Damped linear harmonic oscillator
($\lambda_1\neq 0,~\lambda_2 = 0)$}
Considering  $\lambda_1$ as  arbitrary and $\lambda_2=0$ in Eq. (\ref{3.15}), we obtain from (\ref{1.1}),  the equation of a damped harmonic oscillator,
\begin{equation}
\ddot{x}+k\dot{x}+\lambda_1 x =0.
\label{3.20}
\end{equation}
The invariance property of this equation has been discussed in detail in
Ref.\onlinecite{cervil:1984a}. Following our procedure we find Eq. (\ref{3.20}) to be
invariant under the infinitesimal transformation,
\begin{eqnarray}
\xi  &=& a_1+a_2\sin2\beta t-a_3\cos2\beta t
+e^{\alpha t}(b_1\cos\beta t+b_2\sin\beta t)x, \nonumber\\
\eta &=& \left[b_3+a_2(\beta\cos2\beta t-\alpha\sin2\beta t)
 +a_3(\beta\sin 2\beta t - \alpha\cos 2\beta t)\right]x
 \nonumber\\
 & & \left.- \left[b_1e^{\alpha t}(\alpha\cos\beta t \right.
 +\beta\sin\beta
t)+b_2e^{\alpha t} (\alpha\sin\beta t -\beta\cos\beta t) \right]x^2
\nonumber\\
 & & + e^{-\alpha t}(d_1\cos\beta t+d_2\sin\beta t),
\label{3.21}
\end{eqnarray}
where $\alpha=\frac{k}{2}$, $\beta=
\frac{1}{2}(4\lambda_1-k^2)^\frac{1}{2}$ and
$a_i,b_i,d_j,\; i=1,2,3,\;j=1,2$, are arbitrary constants.

The $sl(3,R)$ symmetry generators \cite{cervil:1984a} are
\begin{eqnarray}
 X_1 & = & \frac{\partial}{\partial t},\;\;
X_2 = \sin2\beta t\frac{\partial}{\partial t}+x(\beta\cos2\beta t-\alpha
\sin2\beta t)
\frac{\partial}{\partial x},\nonumber\\
 X_3 & = & -\cos2\beta t\frac{\partial}{\partial t}+x(\beta\sin2\beta t-
\alpha\cos2\beta t)
\frac{\partial}{\partial x}, \nonumber\\
 X_4 & = & x e^{\alpha t}\left(\cos\beta t\frac{\partial}{\partial
t}-x(\alpha\cos\beta t +\beta\sin\beta t)
\frac{\partial}{\partial x}\right),\nonumber\\
 X_5 & = & x e^{\alpha t}\left(\sin\beta t\frac{\partial}{\partial
t}+x(\beta\cos\beta t -\alpha\sin\beta t)
\frac{\partial}{\partial x}\right),\nonumber\\
 X_6 & = & x\frac{\partial}{\partial x},\;\;
 X_7 = e^{-\alpha t}\cos\beta t\frac{\partial}{\partial x},\;\;
 X_8 = e^{-\alpha t}\sin\beta t\frac{\partial}{\partial x}.
\label{3.22}
\end{eqnarray}
In the above derivation, $4\lambda_1 \le k^2 $. Then for the choice
$4\lambda_1 = k^2$, we have the following infinitesimal symmetries:
\begin{eqnarray}
\xi = &&a_1+a_2 t+a_3 t^2
+(b_1+(b_2+b_3t)e^{-\frac{k}{2}t})x,\label{3.21a} \\
\eta = && -\left(\frac{3}{2}(b_2+b_3t)-b_3\right)
x^2e^{-\frac{k}{2}t}
+\left(c_1+(a_3-\frac{k}{2}a_2)t-\frac{k}{2}a_3t^2\right)x
+(d_1+d_2t)e^{\frac{-k}{2}t}.
\nonumber
\end{eqnarray}
The respective infinitesimal vector fields forming an $sl(3,R)$ algebra are
\begin{eqnarray}
 X_1 & = & \frac{\partial}{\partial t},\;\;
X_2 =  t\frac{\partial}{\partial t}-\frac{k}{2}tx
\frac{\partial}{\partial x},\;\;
 X_3  =   t^2\frac{\partial}{\partial t}+xt(1-\frac{k}{2} t)
\frac{\partial}{\partial x}, \nonumber\\
 X_4 & = & x e^{kt}\frac{\partial}{\partial t},\;\;
 X_5  =  x e^{-\frac{k}{2}t}\left(\frac{\partial}{\partial t}
 -\frac{3}{2}kx\frac{\partial}{\partial x}\right),\;\;
 X_6  =  xe^{-\frac{k}{2}t}\left(t\frac{\partial}{\partial t}
 -\frac{3}{2}kx\frac{\partial}{\partial x}\right),\nonumber\\
 X_7 & = & x\frac{\partial}{\partial x},\;\;
 X_8 = e^{-\frac{k}{2}t}\frac{\partial}{\partial x}.
\label{3.21b}
\end{eqnarray}
Finally the choice $4\lambda_1> k^2$ leads to the following infinitesimal
symmetries:
\begin{eqnarray}
\xi  & = & a_1+\frac{a_2}{\beta}e^{\beta t}- \frac{a_3}{\beta}e^{-\beta t}
+(b_1e^{(\alpha+\beta)t}+b_2e^{(\alpha-\beta)t})x, \nonumber\\
\eta & = & \left((\beta-\alpha)b_1e^{(\alpha+\beta)t}
-(\beta+\alpha)b_2e^{(\alpha-\beta)t}\right)x^2
+d_1e^{(-\alpha+\beta)t}
\nonumber\\
& & +\left(c_1+\frac{1}{2}(1-\frac{2\alpha }{\beta})a_2 e^{\beta t}
+\frac{1}{2}(1+\frac{2\alpha}{\beta})a_3 e^{-\beta t}\right)x
+d_2e^{-(\alpha+\beta)t},
\label{3.21c}
\end{eqnarray}
where $\alpha = \frac{k}{2},\;
\beta = \frac{1}{2}(k^2-4\lambda_1)^{\frac{1}{2}}$. The associated vector
fields  forming an $sl(3,R)$ algebra are
\begin{eqnarray}
 X_1 & = & \frac{\partial}{\partial t},\;\;
X_2 =  e^{\beta t}\left(\frac{1}{\beta}\frac{\partial}{\partial t}
+\frac{1}{2}(1-\frac{2\alpha}{\beta})x\frac{\partial}{\partial x}\right),\nonumber\\
 X_3  & = &  e^{-\beta t}\left(-\frac{1}{\beta}\frac{\partial}{\partial t}
+\frac{1}{2}(1+\frac{2\alpha}{\beta})x\frac{\partial}{\partial x}\right), \;\;
 X_4  =  e^{(\alpha+\beta) t}\left(x \frac{\partial}{\partial t}
 +(\beta-\alpha)x^2\frac{\partial}{\partial x}\right),\label{3.21d}\\
 X_5 & = & e^{(\alpha-\beta) t}\left(x \frac{\partial}{\partial t}
 -(\alpha+\beta)x^2\frac{\partial}{\partial x}\right),\;\;
 X_6  =  x\frac{\partial}{\partial x},\;\;
 X_7 = e^{(-\alpha+\beta)t}\frac{\partial}{\partial x},\;\;
 X_8 = e^{-(\alpha+\beta)t}\frac{\partial}{\partial x}.
\nonumber
\end{eqnarray}

It is well known that equation (\ref{3.20}) admits a time dependent Hamiltonian
of the form $H=\frac{p^2}{2m}e^{-kt}+\frac{\lambda_1}{2}x^2e^{kt}$.  However,
very recently we have proved that the system (\ref{3.20}) also admits a conservative
Hamiltonian description \cite{Sekar:2007} for all the values of $k$ and
$\lambda_1$. The explicit form of the Hamiltonian can be deduced from
equation (71) of paper I by fixing $\alpha=k$,
$\beta=\lambda_1$ and $q=1$ in it.

\subsubsection{\bf Falling particle in a viscous medium
$(\lambda_2 \ne 0,~ \lambda_1 = 0)$}
Choosing  $\lambda_1=0$ and $\lambda_2$ to be  arbitrary in Eq. (\ref{3.15}),
we obtain from Eq. (\ref{1.1})
\begin{equation}
\ddot{x}+k\dot{x}+\lambda_2=0,
\label{3.24}
\end{equation}
which is the equation of a falling particle in a viscous medium.
Eq.~(\ref{3.24}) can be transformed into (\ref{3.16}) through the
transformation $x \rightarrow x' = x + (\frac{\lambda_2}{k})t $, and so its infinitesimal
symmetries and generators can be derived from (\ref{3.17}) and
(\ref{3.18}) by appropriately replacing the variable $x$.

\subsubsection{\bf Displaced damped harmonic oscillator
$(\lambda_1\ne \lambda_2 \ne 0)$}
Let  $\lambda_1\neq \lambda_2\neq 0$ in Eq. (\ref{3.15}).
Then Eq. (\ref{1.1}) becomes
\begin{eqnarray}
\ddot{x}+k\dot{x}+\lambda_1x +\lambda_2=0.
\label{3.28}
\end{eqnarray}
Here too one can transform Eq. (\ref{3.28}) to the
damped harmonic oscillator (\ref{3.20}) through the transformation
$x \rightarrow x' = x +\frac{\lambda_2}{\lambda_1}$ and so the invariance
properties can be analyzed from the infinitesimal symmetries of the
damped harmonic oscillator.
\section{\bf Maximal Lie point symmetries of Li\'enard  type systems: Nonlinear ODES }
\subsection{\bf Modified Emden Equations: $f = kx$}
From the above study,  we find that the choices $f = 0$ and $f = \mbox{constant} $
lead to {\it linear} ODEs only. Thus to generate nonlinear
ODEs  which admit maximum number of symmetry generators one needs to consider
linear/higher degree polynomial/nonpolynomial forms for the function $f(x)$.
To begin with we consider $f$ to be a linear function of $x$, say $f=kx$,
where $k$ is a constant. As a consequence we get (from Eq. (\ref{2.16}))
\begin{eqnarray}
g=\frac{1}{9} k^2 x^3 +\lambda_1x+\lambda_2,
\label{3.29}
\end{eqnarray}
where $\lambda_1$ and $\lambda_2$ are arbitrary constants. Substituting the
explicit forms of $f$ and $g$ into Eq. (\ref{2.15}) and solving it consistently
we obtain the symmetry generators of Eq.~(\ref{1.1}). For this purpose,
we consider the following four cases, $(i)~\lambda_1 = \lambda_2
= 0$, $(ii)~\lambda_1 \ne 0,~\lambda_2 = 0$, $(iii)~\lambda_2 \ne 0,~\lambda_1 = 0
$ and $(iv)~\lambda_1 \ne \lambda_2 \ne 0$, separately.

\subsubsection{\bf Modified Emden equation: $\lambda_1 = \lambda_2 = 0$}
For $\lambda_1=\lambda_2=0$, we obtain from Eq. (\ref{1.1}) the
following nonlinear ODE
\begin{equation}
\ddot{x}+kx\dot{x}+\frac{k^2}{9}x^3=0,
\label{3.30}
\end{equation}
whose linearization, invariance and integrability properties
have been widely discussed in the recent literature
\cite{mahomed:1985,sarlet:1987,Duarte:1987a,leach:88,nls1,Sekar:2004a}. From
our above analysis, Eq. (\ref{3.30}) is found to be invariant under the
Lie point symmetries
\begin{eqnarray}
\xi & = & a_1+a_2t+(c_2+\frac{k}{2}d_1)t^2-\frac{k}{6}d_2t^3+
\Biggr [b_1+b_2t-\frac{k}{6}(c_1+a_2)t^2
 \nonumber\\
& &
-\frac{k}{6}(c_2+\frac{k}{3}d_1)t^3+\frac{k^2}{36}d_2t^4\Biggr]x,\nonumber\\
\eta  & = & d_1+d_2t+(c_1+c_2 t-\frac{k}{2}d_2 t^2)x+
\left[b_2-\frac{k}{3}(c_1+a_2)t-\frac{k}{2}(c_2+\frac{k}{3}d_1)t^2
 +\frac{k^2}{9}d_2t^3
\right]x^2\nonumber\\
& &
-\frac{k}{3}\left[b_1+b_2t-\frac{k}{6}(c_1+a_2)t^2
- \frac{k}{6}(c_2+\frac{k}{3}d_1)t^3
 +\frac{k^2}{36}d_2t^4\right]x^3,
\label{3.31}
\end{eqnarray}
where $a_i$, $b_i$, $c_i$ and $d_i$, $i=1,2$, are arbitrary constants.
These infinitesimal symmetries lead to the symmetry generators
\begin{eqnarray}
X_1 & = & \frac{\partial}{\partial t},\;
X_2 = t(1-\frac{k}{6}xt)\frac{\partial}{\partial t}-
\frac{k}{3}x^2t(1-\frac{k}{6}xt)\frac{\partial}{\partial x},\;
X_3 = x\frac{\partial}{\partial t}-\frac{k}{3}x^3
\frac{\partial}{\partial x},
\nonumber\\
 X_4 & = & xt\frac{\partial}{\partial t}
+x^2(1-\frac{k}{3}xt)\frac{\partial}{\partial x},\;
X_5 = -\frac{k}{6}xt^2\frac{\partial}{\partial t}
+x(1-\frac{k}{3}xt+\frac{k^2}{18}x^2t^2)
\frac{\partial}{\partial x},
\nonumber\\
X_6 & = & t^2(1-\frac{k}{6}xt)\frac{\partial}{\partial t}+xt(1-
\frac{k}{2}xt+\frac{k^2}{18}x^2t^2)\frac{\partial}{\partial x},
\nonumber
\end{eqnarray}
\begin{eqnarray}
 X_7 & = & \frac{k}{2}t^2(1-\frac{k}{9}xt)
\frac{\partial}{\partial t}+(1-\frac{k^2}{6}t^2x^2+\frac{k^3}{54}
t^3x^3)\frac{\partial}{\partial x},
\nonumber\\
X_8 & = & -\frac{k}{6}t^3(1-\frac{k}{6}xt)
\frac{\partial}{\partial t}+t(1-\frac{k}{2}xt+\frac{k^2}{9}x^2t^2
-\frac{k^3}{108}x^3t^3)\frac{\partial}{\partial x},
\label{3.32}
\end{eqnarray}
which also satisfy an $sl(3,R)$ algebra \cite{mahomed:1985}.
Eq. (\ref{3.30}) has the general solution
\begin{align}
x(t) = \frac{t+I_1}{\frac{k}{6}t^2+\frac{I_1k}{3}t+I_2},
\label{3.33}
\end{align}
where $I_1$ and $I_2$ are two integrals of motion with the explicit forms
\begin{align}
I_1  =  -t+\frac{x}{\frac{k}{3}x^2+\dot{x}},\quad
I_2  =  \frac{k}{6}t^2+\frac{1-\frac{k}{3}tx}{\frac{k}{3}x^2+\dot{x}}.
\label{3.34}
\end{align}
\subsubsection{\bf Modified Emden
equation with linear term $(\lambda_1 \neq 0,~ \lambda_2 = 0)$}
Choosing  $\lambda_1 $ as arbitrary and $\lambda_2= 0$ in Eq. (\ref{3.29}), we obtain
\begin{eqnarray}
g=\frac{k^2}{9}x^3+\lambda_1 x.
\label{3.35}
\end{eqnarray}
As a result,  we have a nonlinear ODE of the form
\begin{eqnarray}
\ddot{x}+kx\dot{x}+\frac{k^2}{9}x^3+\lambda_1 x=0, \quad \lambda_1 \neq 0.
\label{3.36}
\end{eqnarray}
Using the forms of  $f$ and $g$ in Eq. (\ref{2.15}),
we obtain the following set of coupled ODEs
for the four arbitrary functions $a$, $b$, $c$ and $d$:
\begin{eqnarray}
& & 3 \ddot{b}+3 \lambda_1 b+ k\dot{a}+kc = 0,\quad
2\dot{c}-\ddot{a}+kd = 0,\quad
\ddot{c}+2 \lambda_1 \dot{a}+ k\dot{d} = 0,
\nonumber\\
& & 3\dddot{b}+3\lambda_1\dot b+3k\dot{c}+k^2d = 0,
\quad \ddot{d}+\lambda_1 d = 0.
\label{3.37}
\end{eqnarray}
Note that the fourth equation in (\ref{3.37}) can be obtained
from first and second equations and one has effectively only
four equations for the four
unknowns. Solving equations (\ref{3.37}) we find that eight parameter Lie point
symmetry groups exist for both the choices $(i)\;\lambda_1>0$ and $(ii)\;\lambda_1<0$.
In the first case the infinitesimal symmetries read
\begin {eqnarray}
\xi  & = & a_1-\frac{c_2}{2\alpha^2}\sin 2\alpha t-\frac{c_3}{2\alpha^2}\cos 2\alpha t-\frac{k}{3\alpha^2}d_1
\sin\alpha t -\frac{k}{3\alpha^2}d_2\cos\alpha t+(b_1\sin \alpha t
\nonumber\\
 & & +b_2\cos \alpha t+\frac{k}{6\alpha^2}a_3
\cos 2 \alpha t +\frac{k}{6\alpha^2}a_2\sin 2\alpha
t-\frac{kc_1}{3\alpha^2})x,\nonumber\\
\eta  & = & (c_1+\frac{a_3}{2}\cos2\alpha t+\frac{a_2}{2}\sin2\alpha t
+\frac{k}{3\alpha}d_1\cos\alpha t-
\frac{k}{3\alpha}d_2\sin\alpha t)x
\nonumber\\
& & +(b_1 \alpha \cos\alpha t- b_2 \alpha  \sin\alpha t-\frac{k}{3\alpha}a_3
 \sin2\alpha t+\frac{k}{3\alpha}a_2\cos2\alpha t)x^2
  \nonumber\\
 & &   -\frac{1}{3}(b_1\sin\alpha t+ b_2 \cos\alpha t
 +\frac{k}{6\alpha^2}a_3\cos2\alpha t
 +\frac{k}{6\alpha^2}a_2\sin2\alpha t-\frac{kc_1}{3\alpha^2})x^3
 \nonumber\\
 & & +d_1\sin\alpha t +d_2\cos\alpha t,
\label{3.338}
\end{eqnarray}
where $a_1,b_i,c_j,d_i $, $i=1,2,\;j=1,2,3$, are arbitrary constants and
$\alpha = \sqrt{\lambda_1}$.

The corresponding symmetry generators are
\begin{eqnarray}
X_1 & = & \frac{\partial}{\partial t}, \;
X_2  = (\sin 2\alpha t+\frac{k}{3\alpha}x\cos 2\alpha t)
\frac{\partial}{\partial t}+x(\alpha \cos 2\alpha t-\frac{2k}
{3}x\sin 2\alpha t\nonumber\\
&&-\frac{k^2}{9\alpha}x^2 \cos2\alpha t)
\frac{\partial}{\partial x},\nonumber\\
 X_3  & = & (\cos2\alpha t-\frac{k}{3\alpha}x\sin2\alpha t)
\frac{\partial}{\partial t}-x(\alpha\sin2\alpha t+\frac{2k}{3}x
\cos2\alpha t\nonumber\\
&&-\frac{k^2}{9\alpha}x^2\sin2\alpha t)
\frac{\partial}{\partial x},\nonumber\\
 X_4 & = & x\sin\alpha t\frac{\partial}{\partial t}+
x^2(\alpha\cos\alpha t-\frac{k}{3}x\sin\alpha t)
\frac{\partial}{\partial x},\nonumber \\
 X_5 & = & x\cos\alpha t \frac{\partial}{\partial t}-x^2
(\alpha\sin\alpha t+\frac{k}{3}x\cos\alpha t)
\frac{\partial}{\partial x},\nonumber\\
X_6 & = & x\frac{\partial}{\partial
t}-(\frac{3\alpha^2 x}{k}+\frac{1}{3}x^3)
\frac{\partial}{\partial x}, \;\;
 X_7 =  -\frac{k}{3\alpha^2}\sin\alpha t
\frac{\partial}{\partial t}+(\sin\alpha t+\frac{k}
{3\alpha}x\cos\alpha t)\frac{\partial}{\partial x},\nonumber\\
 X_8 & = & -\frac{k}{3\alpha^2}\cos\alpha t
\frac{\partial}{\partial t}+(\cos\alpha t-\frac{k}
{3\alpha}x\sin\alpha t)\frac{\partial}{\partial x}.
\label{3.39}
\end{eqnarray}
The vector fields (\ref{3.39}) can be shown to form an $sl(3,R)$ algebra.  For the
case (ii) the infinitesimal symmetries involve exponential functions of
$t$, that is
\begin{eqnarray}
\xi & = & a_1+a_2e^{2\alpha t}+a_3e^{-2\alpha t}+{\displaystyle \frac{k}{3\alpha^2}d_1}e^{\alpha t}
+{\displaystyle \frac{k}{3\alpha^2}d_2}e^{-\alpha t}
+[{\displaystyle \frac{k}{3\alpha^2}c_1}+{\displaystyle b_1}e^{\alpha t}
+{\displaystyle b_2}e^{-\alpha t}\nonumber\\
& & -{\displaystyle \frac{k}{3\alpha}}
(a_2e^{2 \alpha t}-a_3e^{-2\alpha t})]x,\label{3.338a}\\
 \eta  & = & -{\displaystyle \frac{1}{3}\Big[\frac{kc_1}{3\alpha ^2}}+{\displaystyle b_1}e^{\alpha t}+{\displaystyle b_2}e^{-\alpha t}-{\displaystyle \frac{k}{3\alpha}}(a_2e^{2\alpha t}
-a_3e^{-2\alpha t})\Big]x^3
+[{\displaystyle \alpha b_1} e^{\alpha t}
-{\displaystyle \alpha b_2} e^{-\alpha t}
-{\displaystyle \frac{2k}{3}}(a_2e^{2\alpha t}
\nonumber\\
& & + a_3e^{-2\alpha t})]x^2
+[ c_1+\alpha(a_2e^{2\alpha t}-{\displaystyle a_3e^{-2\alpha t})}-{\displaystyle \frac{k}{3\alpha}}(d_1e^{\alpha t}
-d_2e^{-\alpha t})]x
%\nonumber\\
%& &
+{\displaystyle d_1}e^{\alpha t}+{\displaystyle d_2}e^{-\alpha t},
\nonumber
\end{eqnarray}
where $a_i$, $c_1$,$b_j, d_j$, $i = 1, 2, 3$, $j = 1, 2$, are arbitrary constants.

The corresponding symmetry generators, which also satisfy an $sl(3,R)$ algebra, are
\begin{eqnarray}
& &X_1 =  \frac{\partial}{\partial t},\;\;\;
X_2 =  e^{2\alpha t}\Big[\Big(1-\frac{k}{3\alpha}x\Big)\frac{\partial}{\partial t}+
\Big[\Big(\frac{k}{9\alpha}x^3-\frac{2k}{3}x^2
+\alpha x\Big)\frac{\partial}{\partial x}\Big],\nonumber\\
& &X_3   =  e^{-2\alpha t}\Big[\Big(1+\frac{k}{3\alpha}x\Big)\frac{\partial}{\partial t}-
\Big(\frac{k}{9\alpha}x^3+\frac{2k}{3}x^2
+\alpha x\Big)\frac{\partial}{\partial x}\Big],\nonumber\\
& &X_4  = x\frac{\partial}{\partial t}-
\Big(\frac{k}{3}x^3-\frac{3\alpha^2}{k}x\Big)\frac{\partial}{\partial x},\qquad \qquad \quad
 X_5  =  e^{\alpha t}\Big[x\frac{\partial}{\partial t}-\Big(\frac{k}{3}x^3-\alpha x^2\Big)\frac{\partial}{\partial x}\Big],\nonumber\\
& &X_6  =  e^{-\alpha t}\Big[x\frac{\partial}{\partial t}-\Big(\frac{k}{3}x^3+\alpha x^2\Big)\frac{\partial}{\partial x}\Big],\qquad \quad
X_7  = e^{\alpha t}\Big[\frac{\partial}{\partial t}-
\Big(\alpha x-\frac{3\alpha^2}{k}\Big)\frac{\partial}{\partial x}\Big],\nonumber\\
& &X_8  =  e^{-\alpha t}\Big[\frac{\partial}{\partial t}+\Big(\alpha x+ \frac{3\alpha^2}{k}\Big)\frac{\partial}{\partial x}\Big].
\label{3.39a}
\end{eqnarray}

The nonlinear oscillator equation (\ref{3.36}) admits several unusual dynamical
properties.  In fact the dissipative system (\ref{3.36}) admits a
time independent first integral and a conservative
Hamiltonian description\cite{vkc1:2005} for all values of $k$ and $\lambda$, that is
\begin{align}
H=\frac{9\lambda ^2}{2k^2}
\bigg(2-2(1-\frac{2kp}{3\lambda })^{\frac{1}{2}}
+\frac{k^2x^2}{9\lambda }-\frac{2kp}{3\lambda }-\frac{2k^3x^2p}{27\lambda ^2}
\bigg),
\label{lam123}
\end{align}
where
\begin{eqnarray}
p=\frac{ \partial L}{\partial \dot{x}}=-\frac{27\lambda ^3}{2k}
\bigg(\frac{1}{(k\dot{x}+\frac{k^2}{3}x^2+3\lambda )^2}\bigg)
+\frac{3\lambda }{2k}. \label{lam122}
\end{eqnarray}

For $\lambda > 0$, the system (\ref{3.36}) admits explicit amplitude independent sinusoidal periodic
solution of the form
\begin{align}
x(t) = \frac{A\sin(\omega t+\delta)}{1-\frac{k}{3\omega}A\cos(\omega t+\delta)},
\quad 0 \le A \le \frac{\omega}{k}, \;\; \omega = \sqrt{\lambda}
\label{lam119a}
\end{align}
and when $\lambda \le 0$, the solution turn out to be a front like
one\cite{vkc1:2005}, that is
\begin{align}
x(t)=\frac{3\beta(I_1e^{2\beta t}-1)}
{kI_1I_2e^{\beta t}+k(1+I_1e^{\beta t})},
\label{lam119b}
\end{align}
where $\beta=\sqrt{|\lambda |}\;$ and $A, \delta, I_1$, $I_2$ are constants.

\subsubsection{\bf Modified
Emden equation with constant external forcing $(\lambda_2 \neq 0,~\lambda_1 = 0)$}

In Eq. (\ref{3.29}) let us put  $\lambda_1 = 0$ and $\lambda_2 = \mbox{arbitrary}$
so that Eq. (\ref{1.1}) becomes
\begin{eqnarray}
\ddot{x}+kx\dot{x}+\frac{k^2}{9} x^3+\lambda_2=0.
\label{3.41}
\end{eqnarray}

Substituting $f=kx$ and $g=\frac{k^2}{9}x^3+\lambda_2$ in (\ref{2.15}), we obtain the following determining equations for the arbitrary functions $a$, $b$, $c$ and $d$:
\begin{eqnarray}
& & 3\ddot{b}+k\dot{a}+kc  = 0,\quad
2\dot{c}-\ddot{a}+3\lambda_2 b +kd  =  0,\quad
\ddot{c}+k\dot{d}  =  0, \nonumber\\
& &\ddot{d}-\lambda_2c+2\lambda_2\dot{a}  =  0.
\label{addcon1}
\end{eqnarray}
Solving them consistently one obtains
\begin{eqnarray}
a(t) & = & a_1-k(m_1 b_2e^{p_1t}+m_2 b_3e^{p_2t}+m_2 b_4e^{p_3t}
+m_1 b_5e^{p_4t}+m_1 b_6e^{p_5t}+m_2 b_7e^{p_6t}),
\nonumber \\
b(t) & = & b_1+\frac{b_2}{p_1}e^{p_1t}+\frac{b_3}{p_2}e^{p_2t}
+\frac{b_4}{p_3}e^{p_3t}+\frac{b_5}{p_4}e^{p_4t}+\frac{b_6}{p_5}e^{p_5t}
+\frac{b_7}{p_6}e^{p_6t},
\nonumber \\
c(t) & = & -k(m_2p_1b_2e^{p_1t}+m_1p_2b_3e^{p_2t}
+m_1p_3b_4e^{p_3t}+m_2p_4b_5e^{p_4t}
+m_2p_5b_6e^{p_5t}+m_1p_6b_7e^{p_6t}),
\nonumber \\
d(t) & = & -\frac{3\lambda_2}{k}b_1+r_1 b_2e^{p_1t}
+r_2 b_3e^{p_2t}+r_3 b_4e^{p_3t}
+r_4 b_5e^{p_4t}+r_5 b_6e^{p_5t}
+r_6 b_7e^{p_6t}.
\label{addcon2}
\end{eqnarray}
The above derivation is presented in Appendix A.
Here $a_1$ and $b_i$, $i=1,\ldots,7$, are integration constants
which are also the eight symmetry parameters.
The constants $p_i$ and $\gamma_i,$ where $i=1,\ldots,6$, are introduced for simplicity which depend on the
system parameters and can be fixed from the relations (\ref{app116}),
 and (\ref{app121}).  Finally, the constants $m_1$ and $m_2$ are
defined in the expression (\ref{app120}).

Eq. (\ref{addcon2}) provides us the infinitesimal symmetries of the form
\begin{eqnarray}
\xi & = & a_1 - k(m_1b_2e^{p_1t} + m_2b_3e^{p_2t}
 + m_2b_4e^{p_3t} + m_1b_5e^{p_4t} + m_1b_6e^{p_5t}
+ m_2b_7e^{p_6t})
\nonumber\\
& & +\left(b_1+\frac{b_2}{p_1}e^{p_1t}+\frac{b_3}{p_2}e^{p_2t}
+\frac{b_4}{p_3}e^{p_3t}
+\frac{b_5}{p_4}e^{p_4t}+\frac{b_6}{p_5}e^{p_5t}
+\frac{b_7}{p_6}e^{p_6t}\right)x,
\nonumber \\
\eta & = & (b_2e^{p_1t}+b_3e^{p_2t}+b_4e^{p_3t}+b_5e^{p_4t}+b_6e^{p_5t}
+b_7e^{p_6t})x^2 - \frac{k}{3}\left(b_1+\frac{b_2}{p_1}e^{p_1t}
\right. \nonumber \\
& & \left.  +\frac{b_3}{p_2}e^{p_2t}+\frac{b_4}{p_3}e^{p_3t}
+\frac{b_5}{p_4}e^{p_4t}+\frac{b_6}{p_5}e^{p_5t}
+\frac{b_7}{p_6}e^{p_6t}\right)x^3
-\left(m_2p_1b_2e^{p_1t}
\right. \nonumber \\
& & \left.
 +m_1p_2b_3e^{p_2t}+m_1p_3b_4e^{p_3t}+m_2p_4b_5e^{p_4t}+m_2p_5b_6e^{p_5t}
 +m_1p_6b_7e^{p_6t}\right)x
\nonumber\\
& & -\frac{3\lambda_2}{k}b_1+r_1 b_2e^{p_1t}+r_2 b_3e^{p_2t}+r_3 b_4e^{p_3t}+r_4 b_5e^{p_4t}
+r_5 b_6e^{p_5t}+r_6 b_7e^{p_6t}.
\label{addcon3}
\end{eqnarray}

The associated infinitesimal generators are
\begin{eqnarray}
X_1 & = & \frac{\partial}{\partial t},\;\;\;
X_2 = x\frac{\partial}{\partial t}-\left(\frac{3}{k}\lambda_2
 +\frac{k}{3}x^3\right)\frac{\partial}{\partial x},
\nonumber\\
X_3 & = & e^{p_1t}\Biggr[\left(-km_1+\frac{x}{p_1}\right)
\frac{\partial}{\partial t}
+\left(r_1-m_2p_1x+x^2-\frac{k}{3p_1}x^3\right)
\frac{\partial}{\partial x}\Biggr],
\nonumber\\
X_4 & = &  e^{p_2t}\Biggr[\left(-km_2+\frac{x}{p_2}\right)
\frac{\partial}{\partial t}
+\left(r_2-m_1p_2x+x^2-\frac{k}{3p_2}x^3\right)
\frac{\partial}{\partial x}\Biggr],
\nonumber\\
 X_5 & = & e^{p_3t}\Biggr[\left(-km_2+\frac{x}{p_3}\right)
\frac{\partial}{\partial t}
+\left(r_3-m_1p_3x+x^2-\frac{k}{3p_3}x^3\right)
\frac{\partial}{\partial x}\Biggr],\nonumber\\
 X_6 & = & e^{p_4t}\Biggr[\left(-km_1+\frac{x}{p_4}\right)
\frac{\partial}{\partial t}
+\left(r_4-m_2p_4x+x^2-\frac{k}{3p_4}x^3\right)
\frac{\partial}{\partial x}\Biggr],
\nonumber\\
X_7 & = & e^{p_5t}\Biggr[\left(-km_1+\frac{x}{p_5}\right)
\frac{\partial}{\partial t}
+\left(r_5-m_2p_5x+x^2-\frac{k}{3p_5}x^3\right)
\frac{\partial}{\partial x}\Biggr],
\nonumber\\
  X_8 & = & e^{p_6t}\Biggr[\left(-km_2+\frac{x}{p_6}\right)
\frac{\partial}{\partial t}
+\left(r_6-m_1p_6x+x^2-\frac{k}{3p_6}x^3\right)
\frac{\partial}{\partial x}\Biggr].
\label{55}
\end{eqnarray}
The general solution of Eq. (\ref{3.41}) can be deduced explicitly
following Ref. \onlinecite{Sekar:2005} as
\begin{align}
x(t) = \frac{-3\alpha}{k}+\frac{1}{3k}\left(
\frac{18\alpha^2(1-I_1e^{\pm i\alpha\sqrt{3}t})}
{3\alpha(1-I_1e^{\pm i\alpha\sqrt{3}t})\mp 2i\alpha \sqrt{3}I_1I_2
e^{\frac{3\alpha+i\alpha \sqrt{3}}{2}t}\pm i \alpha \sqrt{3}(1+I_1e^{\pm i\alpha\sqrt{3}t})}
\right).
\end{align}
where $\alpha$ is solution of the equation $\alpha^3+\lambda_1 \alpha
-\frac{k}{3}\lambda_2 = 0$ and $I_1$ and $I_2$ are two integrals of motion
whose explicit forms have also been reported in Ref.\onlinecite{Sekar:2005}.
\subsubsection{\bf Modified
Emden equation with linear term and  constant external forcing
$(\lambda_1 \neq 0,~ \lambda_2 \neq 0)$}
Finally, we choose $\lambda_1 \neq$ $\lambda_2 \neq 0$ in
Eq. (\ref{3.29}), so that
\begin{equation}
g=\frac{k^2}{9} x^3 +\lambda_1x+\lambda_2.\nonumber
\end{equation}
For this case Eq. (\ref{1.1}) becomes
\begin{equation}
\ddot{x}+kx\dot{x}+\frac{1}{9}k^2 x^3+\lambda_1 x+\lambda_2=0.
\label{3.42}
\end{equation}
The determining equations for the infinitesimal symmetries become
\begin{eqnarray}
& & 3\ddot{b}+k\dot{a}+3\lambda_1 b+kc  =  0,\;\;\;\;
2\dot{c}-\ddot{a}+3\lambda_2 b +kd  =  0,\nonumber\\
& & \ddot{c}+2\lambda_1\dot{a}+k\dot{d}  =  0,,\;\;\;\;
\ddot{d}-\lambda_2c+2\lambda_2\dot{a}+\lambda_1d  =  0.
\label{addcon10}
\end{eqnarray}
Even though Eq.~(\ref{addcon10}) looks similar to (\ref{addcon1}), the
appearance of additional terms  makes it  tedious to solve.
We solve Eq.~(\ref{addcon10}) with the assumption that
$k,\lambda_1,\lambda_2 \neq 0$ to obtain
\begin{eqnarray}
a(t) & = & a_1-\frac{2\lambda_1^2}{k^2\lambda_2}b_1+\beta_1b_2e^{\alpha_1 t}
+\beta_2b_3e^{-\alpha_1t}
+ b_4e^{\alpha_2t}(\beta_3\cos\alpha_3t
+\beta_4\sin\alpha_3 t)\nonumber\\
& &  + b_5e^{\alpha_2t}(\beta_3\sin\alpha_3 t
-\beta_4\cos\alpha_3t)+ b_6e^{-\alpha_2t}(\beta_5\cos\alpha_3t
+\beta_4\sin\alpha_3 t)\nonumber\\
 & &  +b_7e^{-\alpha_2t}(\beta_4\cos\alpha_3t
-\beta_{5}\sin\alpha_3 t), \nonumber\\
b(t) & = & b_1+\frac{b_2}{\alpha_1}e^{\alpha_1 t}
-\frac{b_3}{\alpha_1}e^{-\alpha_1t}
+ \frac{b_4e^{\alpha_2t}}{(\alpha_2^2+\alpha_3^2)}(\alpha_2\cos\alpha_3t
+\alpha_3\sin\alpha_3 t)\nonumber\\
& &
 + \frac{b_5e^{\alpha_2t}}{(\alpha_2^2+\alpha_3^2)}
(\alpha_2\sin\alpha_3t-\alpha_3\cos\alpha_3 t)
+ \frac{b_6e^{-\alpha_2t}}{(\alpha_2^2+\alpha_3^2)}(\alpha_3\sin\alpha_3 t
-\alpha_2\cos\alpha_3t)
 \nonumber\\
& &
+ \frac{b_7e^{-\alpha_2t}}{(\alpha_2^2+\alpha_3^2)}
(\alpha_2\sin\alpha_3t+\alpha_3\cos\alpha_3 t),
\nonumber\\
c(t) & = & -\frac{3\lambda_1}{k}b_1+\gamma_1b_2e^{\alpha_1 t}
+\gamma_2b_3e^{-\alpha_1t}
+ b_4e^{\alpha_2t}(\gamma_3\cos\alpha_3t
+\gamma_4\sin\alpha_3 t)\nonumber\\
& &
 + b_5e^{\alpha_2t}(\gamma_3\sin\alpha_3t-\gamma_4\cos\alpha_3t)
+ b_6e^{-\alpha_2t}(\gamma_5\cos\alpha_3t
+\gamma_6\sin\alpha_3 t)\nonumber\\
& &
 +b_7e^{-\alpha_2t}(\gamma_6\cos\alpha_3t
-\gamma_{5}\sin\alpha_3 t),
\nonumber\\
d(t) & = & -\frac{3\lambda_2}{k}b_1+\delta_1b_2e^{\alpha_1 t}
+\delta_2b_3e^{-\alpha_1t}
+ b_4e^{\alpha_2t}(\delta_3\cos\alpha_3t
+\delta_4\sin\alpha_3 t)\nonumber\\
& &
 + b_5e^{\alpha_2t}(\delta_3\sin\alpha_3t-\delta_4\cos\alpha_3t)
+ b_6e^{-\alpha_2t}(\delta_5\cos\alpha_3t
+\delta_6\sin\alpha_3 t)\nonumber\\
& &
 +b_7e^{-\alpha_2t}(\delta_6\cos\alpha_3t
-\delta_{5}\sin\alpha_3 t).
\label{addcon11}
\end{eqnarray}
Here $a_1$ and $b_i$, $i=1,\ldots,7$, are
the arbitrary
constants arising from the integration of Eq.~(\ref{addcon10}) which fix
the eight symmetry generators.  The other constants are included for
simplicity sake.  A detailed analysis on the above derivation and the relation
between these constants and system parameters are given in Appendix B.

The infinitesimal symmetries turn out to be
\begin{eqnarray}
\xi & = & a_1-\frac{2\lambda_1^2}{k^2\lambda_2}b_1+\beta_1b_2e^{\alpha_1 t}
+\beta_2b_3e^{-\alpha_1t}
+ b_4e^{\alpha_2t}(\beta_3\cos\alpha_3t
+\beta_4\sin\alpha_3 t)+ b_5e^{\alpha_2t}(\beta_3\sin\alpha_3 t
 \nonumber\\
& & -\beta_4\cos\alpha_3t)
+ b_6e^{-\alpha_2t}(\beta_5\cos\alpha_3t
+\beta_4\sin\alpha_3 t)+b_7e^{-\alpha_2t}(\beta_4\cos\alpha_3t
-\beta_{5}\sin\alpha_3 t)
\nonumber\\
& & +b_7e^{-\alpha_2t}(\beta_4\cos\alpha_3t
-\beta_{5}\sin\alpha_3 t)
+\left[b_1+\frac{b_2}{\alpha_1}e^{\alpha_1 t}
-\frac{b_3}{\alpha_1}e^{-\alpha_1t}
+ \frac{b_4e^{\alpha_2t}}{(\alpha_2^2+\alpha_3^2)}(\alpha_2cos\alpha_3t
\right. \nonumber\\
& &
\left. +\alpha_3sin\alpha_3 t)
+ \frac{b_5e^{\alpha_2t}}{(\alpha_2^2+\alpha_3^2)}(\alpha_2\sin\alpha_3t
-\alpha_3\cos\alpha_3 t)+\frac{b_6e^{-\alpha_2t}}{(\alpha_2^2+\alpha_3^2)}
(\alpha_3\sin\alpha_3 t-\alpha_2\cos\alpha_3t)
\right. \nonumber\\
& &
\left.+ \frac{b_7e^{-\alpha_2t}}{(\alpha_2^2+\alpha_3^2)}(\alpha_2\sin\alpha_3t
+\alpha_3\cos\alpha_3 t)\right]x,
\nonumber
\end{eqnarray}
\begin{eqnarray}
\eta & = & \frac{k}{3}\left[b_1+\frac{b_2}{\alpha_1}e^{\alpha_1 t}
-\frac{b_3}{\alpha_1}e^{-\alpha_1t}
+ \frac{b_4e^{\alpha_2t}}{(\alpha_2^2+\alpha_3^2)}(\alpha_2\cos\alpha_3t
+\alpha_3\sin\alpha_3 t)
+\frac{b_5e^{\alpha_2t}}{(\alpha_2^2+\alpha_3^2)}(\alpha_2\sin\alpha_3t
\right. \nonumber \\
& &
\left. -\alpha_3\cos\alpha_3 t)
+\frac{b_6e^{-\alpha_2t}}{(\alpha_2^2+\alpha_3^2)}(\alpha_3\sin\alpha_3 t
  -\alpha_2\cos\alpha_3t)+\frac{b_7e^{-\alpha_2t}}
{(\alpha_2^2+\alpha_3^2)}(\alpha_2\sin\alpha_3t
\right.
\nonumber \\
  & & \left.
  +\alpha_3\cos\alpha_3 t)\right]x^3
 +\Biggr[b_2e^{\alpha_1t}+b_3e^{-\alpha_1t}
+e^{\alpha_2t}(b_4\cos\alpha_3t+b_5\sin\alpha_3t)
+e^{-\alpha_2t}(b_6\cos\alpha_3t
  \nonumber \\
%\nonumber\\
& &
-b_7\sin\alpha_3t)\Biggr]x^2
+\Biggr[-\frac{3\lambda_1}{k}b_1+\gamma_1b_2e^{\alpha_1 t}
+\gamma_2b_3e^{-\alpha_1t}
+ b_4e^{\alpha_2t}(\gamma_3\cos\alpha_3t
+\gamma_4\sin\alpha_3 t)
\nonumber\\
%\end{eqnarray}
%\begin{eqnarray}
& &
 + b_5e^{\alpha_2t}(\gamma_3\sin\alpha_3t-\gamma_4\cos\alpha_3t)
+ b_6e^{-\alpha_2t}(\gamma_5\cos\alpha_3t
+\gamma_6\sin\alpha_3 t)
+b_7e^{-\alpha_2t}(\gamma_6\cos\alpha_3t
\nonumber\\
& & -\gamma_{5}\sin\alpha_3 t)\Biggr]x
-\frac{3\lambda_2}{k}b_1+\delta_1b_2e^{\alpha_1 t}
+\delta_2b_3e^{-\alpha_1t}
+ b_4e^{\alpha_2t}(\delta_3\cos\alpha_3t
+\delta_4\sin\alpha_3 t)
\nonumber\\
& & + b_5e^{\alpha_2t}(\delta_3\sin\alpha_3t-\delta_4\cos\alpha_3t)
+ b_6e^{-\alpha_2t}(\delta_5\cos\alpha_3t
+\delta_6\sin\alpha_3 t)
\nonumber\\
& & +b_7e^{-\alpha_2t}(\delta_6\cos\alpha_3t
-\delta_{5}\sin\alpha_3 t).\label{addcon11a}
\end{eqnarray}

The associated generators constituting a $sl(3,R)$ algebra are
\begin{eqnarray}
X_1 & = & \frac{\partial}{\partial t},\;
X_2 = \left(x-\frac{2\lambda_1^2}{k^2\lambda_2^2}\right)
\frac{\partial}{\partial t}+\left(-\frac{3\lambda_2}{k}-
\frac{3\lambda_1}{k}x+\frac{k}{3}x^3\right)
\frac{\partial}{\partial x},
\nonumber\\
X_3 & = & \left[\left(\beta_1+\frac{x}{\alpha_1}\right)
\frac{\partial}{\partial t}+\left(\delta_1+\gamma_1x+x^2+\frac{k}{3\alpha_1}x^3
\right)\frac{\partial}{\partial x}\right]e^{\alpha_1t},
\nonumber\\
X_4 & = & \left[\left(\beta_2-\frac{x}{\alpha_1}\right)
\frac{\partial}{\partial t}+\left(\delta_2+\gamma_2x+x^2-\frac{k}{3\alpha_1}x^3
\right)\frac{\partial}{\partial x}\right]e^{-\alpha_1t},
\nonumber
\end{eqnarray}
\begin{eqnarray}
X_5 & = & \Biggr[\left(\beta_3\cos\alpha_3t+\beta_4\sin\alpha_3t
+\frac{1}{(\alpha_2^2+\alpha_3^3)}(\alpha_2\cos\alpha_3t
+\alpha_3\sin\alpha_3t)x\right)\frac{\partial}{\partial t}
\nonumber\\
& &
+\left(\delta_3\cos\alpha_3t+\delta_4\sin\alpha_3t
+(\gamma_3\cos\alpha_3t+\gamma_4\sin\alpha_3t)x+\cos\alpha_3t x^2
\right.
\nonumber\\
 & &
\left. +\frac{1}{(\alpha_2^2+\alpha_3^2)}
(\alpha_2\cos\alpha_3t+\alpha_2\sin\alpha_3t)x^3\right)
\frac{\partial}{\partial x}\Biggr]e^{\alpha_2t},
\nonumber\\
X_6 & = & \Biggr[\left(\beta_3\sin\alpha_3t-\beta_4\cos\alpha_3t
+\frac{1}{(\alpha_2^2+\alpha_3^3)}(\alpha_2\sin\alpha_3t
-\alpha_3\cos\alpha_3t)x\right)\frac{\partial}{\partial t}
\nonumber\\
& &
+\Biggr(-\delta_4\cos\alpha_3t+\delta_3\sin\alpha_3t
+(\gamma_3\sin\alpha_3t-\gamma_4\cos\alpha_3t)x+\sin\alpha_3t x^2
\nonumber\\
& &   +\frac{1}{(\alpha_2^2+\alpha_3^2)}
(\alpha_2\sin\alpha_3t-\alpha_3\cos\alpha_3t)x^3\Biggr)
\frac{\partial}{\partial x}\Biggr]e^{\alpha_2t},
\nonumber\\
X_7 & = & \Biggr[\left(\beta_5\cos\alpha_3t+\beta_4\sin\alpha_3t
+\frac{1}{(\alpha_2^2+\alpha_3^3)}(\alpha_3\sin\alpha_3t
-\alpha_2\cos\alpha_3t)x\right)\frac{\partial}{\partial t}
\nonumber\\
& &
+\Biggr(\delta_5\cos\alpha_3t+\delta_6\sin\alpha_3t
+(\gamma_5\cos\alpha_3t+\gamma_6\sin\alpha_3t)x+\cos\alpha_3t x^2
\nonumber\\
& &
 +\frac{1}{(\alpha_2^2+\alpha_3^2)}
(\alpha_3\sin\alpha_3t-\alpha_2\cos\alpha_3t)x^3\Biggr)
\frac{\partial}{\partial x}\Biggr]e^{-\alpha_2t},
\nonumber\\
X_8 & = & \Biggr[\left(\beta_4\cos\alpha_3t
-\beta_{5}\sin\alpha_3t
+\frac{1}{(\alpha_2^2+\alpha_3^3)}(\alpha_2\sin\alpha_3t
+\alpha_3\cos\alpha_3t)x\right)\frac{\partial}{\partial t}
\nonumber\\
& &
+\Biggr(\delta_6\cos\alpha_3t-\delta_{5}\sin\alpha_3t
+(\gamma_6\cos\alpha_3t-\gamma_{5}\sin\alpha_3t)x-\sin\alpha_3t x^2
\nonumber\\
& &
 +\frac{1}{(\alpha_2^2+\alpha_3^2)}
(\alpha_2\sin\alpha_3t+\alpha_3\cos\alpha_3t)x^3\Biggr)
\frac{\partial}{\partial x}\Biggr]e^{-\alpha_2t}.
\label{addcon11aa}
\end{eqnarray}
As far the authors knowledge goes {\it the above symmetry generators are
being reported for the first time}.  Eq. (\ref{3.42})
is also linearizable one as it admits eight parameter Lie
point symmetries.  The general solution\cite{Sekar:2004b} of Eq. (\ref{3.42}) reads as
\begin{align}
x(t) = \frac{-3\alpha}{k}+\frac{6}{k}\left(
\frac{(3\alpha^2+\lambda_1)(1-I_1e^{\pm i\hat{\alpha}t})}
{3(1-I_1e^{\pm \hat{\alpha}t})-2(3\alpha^2+\lambda_1)I_2e^{(-3\alpha\pm
\frac{\hat{\alpha}}{2})t}
\pm \hat{\alpha} (1-I_1e^{\pm \hat{\alpha}t})}
\right)
\label{mgen1}
\end{align}
where
\begin{align}
\alpha^3+\alpha \lambda_1-k\lambda_2 = 0,\;\;and \;\;
\hat{\alpha} = \sqrt{-3\alpha^2-4\lambda_1}.
\end{align}
For more details about the derivation of the solution (\ref{mgen1}) one
may refer to Ref.\onlinecite{Sekar:2004b}.

\subsection{\bf Generalized Modified Emden
Equations: $f = k_1x + k_2$}
Next we consider the case where $f$
as the most general linear function of $x$, that is $f=k_1 x +k_2$,
where $k_1$ and $k_2$ are arbitrary constants.  As a consequence we arrive at
\begin{eqnarray}
g=\frac{k_1^2}{9}x^3 +\frac{k_1k_2}{3}x^2+\lambda_1x+\lambda_2,
\label{3.43}
\end{eqnarray}
where $\lambda_1$ and $\lambda_2$ are  arbitrary constants.

Proceeding in a similar fashion as earlier, we consider the following
four cases:
$(i)~\lambda_1 = \lambda_2 = 0$, $(ii)~\lambda_1
\neq 0,\; \lambda_2 = 0$ $(iii)~\lambda_2
\neq 0, ~\lambda_1 = 0,$ and
$(iv)~\lambda_1 \neq \lambda_2 \neq 0$, separately.  The corresponding
equations of motion, respectively, are
\begin{eqnarray}
&& \ddot{x}+(k_1x+k_2)\dot{x}+\frac{k_1k_2}{3}x^2+\frac{k_1^2}{9}x^3 = 0.
 \label{3.45a} \\
&& \ddot{x}+(k_1x+k_2)\dot{x}+\frac{k_1k_2}{3}x^2+\frac{k_1^2}{9}x^3+
 \lambda_1 x = 0.
 \label{3.45b} \\
 &&\ddot{x}+(k_1x+k_2)\dot{x}+\frac{k_1k_2}{3}x^2+\frac{k_1^2}{9}x^3+
 \lambda_2 = 0.
 \label{3.45c} \\
&& \ddot{x}+(k_1x+k_2)\dot{x}+\frac{k_1k_2}{3}x^2+\frac{k_1^2}{9}x^3+
 \lambda_1 x + \lambda_2 = 0.
 \label{3.45d}
\end{eqnarray}
However, all the above equations can be transformed straightforwardly to the
equations discussed in the previous subsections.  For example with the subtitution,
$X=x+\frac{k_2}{k_1}$, Eq. ~(\ref{3.45a}) can be transformed to the form
\begin{eqnarray}
\ddot{X}+k_1X\dot{X}+\frac{k_1^2}{9}X^3-\frac{k_2^2}{3}X+\frac{2}{9} \frac{k_2^3}{k_1} = 0.
\label{3.45e}
\end{eqnarray}
Equation (\ref{3.45e}) coincides exactly with (\ref{3.42}) by restricting the constants $\lambda_1=-\frac{k_2^2}{3}$ and $\lambda_2=\frac{2}{9} \frac{k_2^3}{k_1}$ in the latter. As a result the infinitesimal symmetries and their
associated
symmetry generators can also be derived from the Eqs.~(\ref{addcon11a}) and
(\ref{addcon11aa}).

Eq.~(\ref{3.45b}) can be cast into the form,
\begin{eqnarray}
\ddot{X}+k_1X\dot{X}+\frac{k_1^2}{9}X^3+\frac{(3\lambda_1-k_2^2)}{3}X+
\frac{(2k_2^3-9k_2\lambda_1)}{9k_1} = 0,
\label{3.45f}
\end{eqnarray}
through the transformation $X=x+\frac{k_2}{k_1}$.  Eq. (\ref{3.45f}) is of the form
(\ref{3.42}).  Therefore, the infinitesimal symmetries and the associated symmetry generators of equation (\ref{3.45f}) can also be derived from the Eqs.~(\ref{addcon11a}) and
(\ref{addcon11aa}) by appropriately fixing the constants.

In a similar way, one can transform Eqs.~(\ref{3.45c}) and (\ref{3.45d}) to the
following form:
\begin{eqnarray}
&\ddot{X}+k_1X\dot{X}+\frac{\displaystyle{k_1^3}}{\displaystyle{9}}X^3
-\displaystyle{\frac{k_2^2}{3}X+\frac{2k_2^3}{9k_1}
+\frac{k_1\lambda_2}{\lambda_1}}
 = 0.
\label{3.45g}\\
&\ddot{X}+k_1X\dot{X}+\frac{\displaystyle{k_1^3}}{\displaystyle{9}}X^3
+\displaystyle{(\lambda_1-\frac{k_2^2}{3})X
+\frac{(2k_2^3+9k_1\lambda_2-9k_2\lambda_1)}{9k_1^2}}
 = 0.
\label{3.45h}
\end{eqnarray}
respectively.  Eqs. (\ref{3.45g}) and (\ref{3.45h}) are also of the form
(\ref{3.42}) and hence we can derive the infinitesimal symmetries directly
from the Eq. (\ref{addcon11a}).

\section{Non-maximal Lie point symmetries for $f_{xx} \neq 0 $}
\subsubsection{\bf General forms of $\mathbf{f(x)}$}
One can consider higher degree polynomials or other forms for $f$ such that $(f_{xx}\neq0)$ and classify the equations and their underlying Lie point symmetries. However, our analysis shows that one can get only lesser number of Lie point symmetries than eight whenever
$f_{xx} \neq0$.
This can be proved in a rather general way as given below.
\par Now we assume $f=k_1x+k_2+l(x)$, where $l(x)$ is an arbitrary function and investigate the outcome when $l(x) \neq 0$. With this choice, $g$ takes the form (vide Eq. (\ref{2.16}))
\begin{equation}
g=\frac{1}{3b}\Big[-(\dot b x^2-2bf_2+cx+d)f_x-(\dot a+\dot b x)+4\dot bf_1-3\ddot b x-2\dot c+\ddot a\Big].\label{3.45i}
\end{equation}
In the above we introduced the notation $f_1=\int{f(x)dx}$ and $f_2$=$\int{f_1(x)dx}$=$\int({\int{f(x)dx})dx}$.
Substituting $g$ and its derivative into Eq. (\ref{2.15}) and expanding the latter we arrive at an expression exclusively in terms of $l(x)$, that is
\begin{eqnarray}
& & \frac{4}{3}{bl\Big(l_4}+\frac{k_2}{2}x^2+\frac{k_1}{6}x^3\Big)^2-\frac{ c^2lx^2}{3b}-\frac{2cdlx}{3b}-\frac{ld^2}{3b}+\frac{4}{3} l\Big(cx+d\Big) \Big(\frac{k_2}{2}x^2+\frac{k_1}{6} x^3+l_4\Big)  \nonumber\\
& & \quad +\frac{c\dot a}{3b} \Big(k_2+k_1 x + l_2\Big)-\frac{2\dot a}{3}\Big(k_2x+\frac{k_1}{2}x^2+l_3\Big)\Big(k_2+k_1x+l_2\Big)-\frac{\dot a}{b} \Big(k_1+l_1\Big)\Big(cx+d\Big)\nonumber\\
& & \quad +2\Big(\frac{k_2}{2} x^2+\frac{k_1}{6} x^3+l_4\Big)\Big(k_1+l_1\Big) \dot a
-\frac{2 \dot a}{3b}\Big(k_2+k_1x+l_2\Big)-\frac{4 \dot b c}{3b}\Big(k_2x+\frac{k_1}{2} x^2+l_3\Big)-\frac{2\dot{a}\ddot{b}x}{b} \nonumber\\
& & \quad +\frac{8}{3} \dot{b}\Big(k_2x+\frac{k_1}{2} x^2+l_3\Big)^2+\frac{\dot {b}}{3b}\Big(k_2+k_1x+l_2\Big)\Big(4 c x+3 d\Big)-\frac{\dot{b}cx^2}{b} \Big(k_1+l_1\Big)-\frac{\dot{b}dx}{b} \Big(k_1+l_1\Big)\nonumber\\
& & \quad -4\Big(\frac{k_2}{2} x^2+\frac{k_1}{6} x^3+l_4\Big)  \Big(k_2+k_1x+l_2\Big)\dot{b}  -\frac{2}{3} x \Big(k_2 x+\frac{k_1}{2} x^2+l_3\Big)\Big(k_2+k_1x+l_2\Big) \dot{b}
-\frac{4\dot{a}\dot{c}}{3b}\nonumber\\
& & \quad + 2x\Big(\frac{k_2}{2}x^2+\frac{k_1}{6} x^3+l_4\Big)\Big(k_1+l_1\Big)\dot b-\frac{2}{3}\dot{b}x^2 l\Big(cx+d\Big)+\frac{4}{3} x^2l\dot{b} \Big(\frac{k_2}{2} x^2+\frac{k_1}{6} x^3+l_4\Big)-\frac{x^2 \dot{b}\ddot{b}}{b}\nonumber\\
& & \quad +\frac{8 \dot{a}\dot{b}}{3b}\Big(k_2x+\frac{k_1}{2}x^2+l_3\Big)-\frac{2\dot{a}\dot {b}}{3b}\Big(k_2+k_1x+l_2\Big)x-\frac{\dot{a}\dot{b}}{b}\Big(k_1+l_1\Big)x^2+\frac{\dot {b}^2}{b}\Big(k_2+k_1 x+l_2\Big)x^2\nonumber\\
& & \quad -\frac{\dot{b}^2x^3}{b}\Big(k_1+l_1\Big)-\frac{l\dot{b}^2x^4}{3b}+\frac{2c\dot c}{3b}-\frac{4}{3}\dot{c}\Big(k_2 x+\frac{k_1}{2}x^2+l_3\Big)+\dot{c}x\Big(k_2+k_1 x+l_2\Big)\nonumber\\
& & \quad +\Big(k_2+k_1 x+l_2\Big)\dot{d}-\frac{c\ddot{a}}{3b}+\frac{2}{3} \Big(k_2x+\frac{k_1}{2}x^2+l_3\Big)\ddot{a}+\frac{2\dot{a}\ddot{a}}{3b}-\frac{d\ddot {b}}{b}-2\ddot{b}\Big(k_2x+k_1x^2+l_3\Big)x\nonumber\\
& & \quad  + x^2 \Big(k_2+k_1 x+l_2\Big) \ddot{b}+\ddot{c}x+\ddot{d}
+\dddot{b}x^2=0.
\label{gen1}
\end{eqnarray}

Again in (\ref{gen1}), for the sake of simplicity, we have used the notation $l_1(x)=\int{l(x)dx}$, $l_2(x)=\int{[\int{l(x)dx}]dx}$ and so on.  It is clear from the
above expression that the first term on the left hand side of Eq. (\ref{gen1}), namely
$\frac{4}{3}{bl\Big(l_4}+\frac{k_2}{2}x^2+\frac{k_1}{6}x^3\Big)^2$ cannot be balanced by
no other term.  This in turn leads us to
\begin{equation}
\frac{4}{3}{bl\Big(l_4}+\frac{k_2}{2}x^2+\frac{k_1}{6}x^3\Big)^2 \Longrightarrow
b =0.
\end{equation}
When $l(x)\neq0$ the above expression clearly indicates that the
symmetry function $b(t)$ should be zero.  Since one of the symmetry functions
becomes zero one can get ultimately lesser Lie point symmetries only. This has in fact been
analyzed in detail in Sec. III of paper I.

On the other
hand restricting $l(x)=0$, that is $f = k_1x+k_2$, then one obtains maximal Lie point symmetries as shown in the earlier sections.

\subsection*{\bf 2. Relationship between the symmetries and the forms
of $\mathbf{f(x)}$ and $\mathbf{g(x)}$}

In our discussions so far, we find that all the examples dealt with in Secs. III and
IV admit eight symmetry generators, and each one of the sets form $sl(3,R)$ algebra. It is a
well known fact that second order ODEs with eight symmetry generators are linearizable
by point transformation\cite{ibra:1999,ma}, which was pointed out by Lie himself \cite{Lie}.  In
fact the theorem that provides criteria for testing the linearization of a scalar
second order ODE $\ddot{x}=f(t,x,\dot{x})$ via a point transformation has the cubic in derivative form \cite{ibra:1999,ma}
\begin{eqnarray}
\ddot{x}+A(t,x)\dot{x}^3+B(t,x)\dot{x}^2+C(t,x)\dot{x}+D(t,x) = 0,
\label{lin1}
\end{eqnarray}
with the coefficients $A,B,C$ and $D$ satisfying  the
following two invariant conditions,
\begin{eqnarray}
&&3A_{tt}+3A_tC-3A_xD+3AC_t+C_{xx}-6AD_x+BC_x-2BB_t-2B_{tx}=0,
\label{lin2}\\
&&6A_tD-3B_xD+3AD_t+B_{tt}-2C_{tx}-3BD_x+3D_{xx}+2CC_x-CB_t=0,
\label{lin3}
\end{eqnarray}
where suffixes refer to partial derivatives. On comparing Eq. (\ref{1.1})
with Eq. (\ref{lin1}), we find that the functions $A(t,x) = B(t,x) = 0$
and $C(t,x)$ and $D(t,x)$ are $f(x)$ and $g(x)$ respectively which are
independent of time. Substituting these forms of $C(t,x)$ and $D(t,x)$
along with $A(t,x) = B(t,x) = 0$ in Eqs. (\ref{lin2}) and (\ref{lin3}),
we obtain
\begin{eqnarray}
f_{xx} = 0, \quad 3g_{xx}-2ff_x = 0.
\label{lin4}
\end{eqnarray}
Integrating Eq. (\ref{lin4}), we obtain
\begin{eqnarray}
f = k_1x+k_2, \quad g = \frac{1}{9}k_1k_2x^3+\frac{1}{3}k_2^2x^2+
\lambda_1 x+\lambda_2,
\label{lin5}
\end{eqnarray}
where $\lambda_1,\lambda_2$ are integration constants. Thus for the function
$f_{xx}\ne0$, one obtains lesser parameter symmetries only . However, our analysis leads
to a more general form from which one can determine systems exhibiting lesser
parameter symmetry groups. Thus this analysis broadens our knowledge on both
integrable
and linearizable equations belonging to Eq. (\ref{1.1}).

Note that the above relation (\ref{lin5}) is the same as the one obtained from the analysis of
the determining equations, Eq. (21) for the Lie symmetries in Sec. II A of paper I for the case
$f_{xx} =0$.

\section{Discussion and Conclusions}

In this paper, we have
studied in some detail the symmetry properties of the Li\'enard type equation
(1), where none of the symmetry functions is equal to zero.  We
have also isolated the class of equations which possess eight Lie point
symmetries.  For the sake of completeness we have also given the solutions
for all the nonlinear ODEs which we have identified as linearizable
in this procedure.   We have also indicated the explicit Hamiltonian forms
for these equations.

A question now naturally arises is about the Noether symmetries of these
equations.   As it is well known, Noether's theorem is based upon the invariance
of action integral and essentially one needs a Lagrangian to start the
analysis.  Since our recent studies show that even dissipative systems
can have time independent Lagrangian \cite{pcsl} and so it is worth exploring
Noether symmetries of these systems as well.  Finally, it is also of importance
to study the detailed symmetry structure when the functions $f$ and $g$
in (1) are also time dependent.  These are being pursued currently.
\section*{Acknowledgments}
One of us (SNP) is grateful to the Centre for Nonlinear Dynamics,
Bharathidasan University, Tiruchirappalli, for warm
hospitality.
 The work of SNP forms part of a
Department of Science and Technology, Government of India
sponsored research project. The work of MS forms part of a
research project sponsored by the National Board for Higher
Mathematics, Government of India.  The work of ML forms part of a
Department of Science and Technology (DST), Ramanna Fellowship and is
also supported by a DST-IRHPA research project.

\appendix
\section{Method of solving the Determining Eqs.~(\ref{addcon1})}
\setcounter{section}{1}
In this section we solve the determining equations (\ref{addcon1})
to obtain the general form of $\xi$ and $\eta$ which leaves
Eq.~(\ref{3.41})
invariant.  The determining equations to be solved  are
%\numparts
\begin{eqnarray}
& &3\ddot{b}+k\dot{a}+kc  =  0,
\label{app101}\\
& &2\dot{c}-\ddot{a}+3\lambda_2 b +kd  = 0,
\label{app102}\\
& &\ddot{c}+k\dot{d}  =  0,
\label{app103}\\
& &\ddot{d}-\lambda_2c+2\lambda_2\dot{a}  =  0.
\label{app104}
\end{eqnarray}
%\endnumparts
In the following, we solve
eqs.~(\ref{app101})-(\ref{app104}) with the assumption that the parameters
$k,\lambda_2 \neq 0$ and deduce the infinitesimal symmetries.

Rewriting (\ref{app101}) we get
\begin{eqnarray}
c = - \frac{3}{k}\ddot{b} - \dot{a}.
\label{app106}
\end{eqnarray}
Substituting (\ref{app106}) into (\ref{app102}) and rearranging,
we obtain an expression for the function $d$ of the form
\begin{eqnarray}
d = \frac{6}{k^2}\dddot{b}+\frac{3}{k}\ddot{a}-\frac{3\lambda_2}{k}b.
\label{app107}
\end{eqnarray}
Now substituting the forms of $c$ and $d$ and their derivatives into
(\ref{app103}) and (\ref{app104}) and simplifying, we get two equations
for the unknowns $a$ and $b$:
\begin{eqnarray}
3\frac{d^4b}{dt^4}+2k\frac{d^3a}{dt^3}-3k\lambda_2\frac{db}{dt} = 0,
\label{app108}\\
6\frac{d^5b}{dt^5}+3k\frac{d^4a}{dt^4}+3k^2\lambda_2\frac{da}{dt} = 0.
\label{app109}
\end{eqnarray}
From Eq.~(\ref{app108}) we have
\begin{eqnarray}
\frac{d^3a}{dt^3} = -\frac{3}{2k}\frac{d^4b}{dt^4} + \frac{3}{2}\lambda_2
\frac{db}{dt}.
\label{app110}
\end{eqnarray}
Substituting  (\ref{app110}) into (\ref{app109}) we get
\begin{eqnarray}
\frac{d^5b}{dt^5}+3k\lambda_2\frac{d^2b}{dt^2}+2k^2\lambda_2\frac{da}{dt}
= 0.
\label{app111}
\end{eqnarray}
Rearranging the above we obtain
\begin{eqnarray}
\frac{da}{dt} = -\frac{1}{2k^2\lambda_2}\frac{d^5b}{dt^5}
-\frac{3}{2k}\frac{d^2b}{dt^2}.
\label{app112}
\end{eqnarray}
On integration (\ref{app112}) yields
\begin{eqnarray}
a(t) = a_1-\frac{1}{2k^2\lambda_2}\frac{d^4b}{dt^4}
-\frac{3}{2k}\frac{db}{dt},
\label{app113}
\end{eqnarray}
where $a_1$ is an integration constant.  Using (\ref{app112}),
Eq.~(\ref{app108}) can be written as
\begin{eqnarray}
\frac{d^7b}{dt^7}+3k^2\lambda_2^2\frac{db}{dt} = 0.
\label{app114}
\end{eqnarray}
A general solution for (\ref{app114}) takes the form
\begin{eqnarray}
b(t) = b_1+\frac{b_2}{p_1}e^{p_1t}+\frac{b_3}{p_2}e^{p_2t}
+\frac{b_4}{p_3}e^{p_3t}+\frac{b_5}{p_4}e^{p_4t}+\frac{b_6}{p_5}e^{p_5t}
+\frac{b_7}{p_6}e^{p_6t},
\label{app115}
\end{eqnarray}
where $b_i's,\;i=1,\ldots,7$, are integration constants and
$p_j's,\;j=1,\ldots,6$, are given by
\begin{eqnarray}
p_1 & = & (-3)^{\frac{1}{6}}(k\lambda_2)^{\frac{1}{3}},\;\;
p_2 = -(-3)^{\frac{1}{6}}(k\lambda_2)^{\frac{1}{3}},\;\;
p_3 = (-1)^{\frac{1}{3}}(-3)^{\frac{1}{6}}(k\lambda_2)^{\frac{1}{3}},\label{app116}\\
p_4 & = & -(-1)^{\frac{1}{3}}(-3)^{\frac{1}{6}}(k\lambda_2)^{\frac{1}{3}},\;\;
p_5  =  (-1)^{\frac{2}{3}}(-3)^{\frac{1}{6}}(k\lambda_2)^{\frac{1}{3}},\;\;
p_6 = -(-1)^{\frac{2}{3}}(-3)^{\frac{1}{6}}(k\lambda_2)^{\frac{1}{3}}.
\nonumber
\end{eqnarray}
Substituting the form of $b$ given by (\ref{app115}),
and its derivatives into (\ref{app113}),
we get
\begin{eqnarray}
a(t) & = & a_1 - \frac{(3+\sqrt{-3})}{2k}b_2e^{p_1t} -
\frac{(3-\sqrt{-3})}{2k}b_3e^{p_2t} - \frac{(3-\sqrt{-3})}{2k}b_4e^{p_3t}
\nonumber \\
& & - \frac{(3+\sqrt{-3})}{2k}b_5e^{p_4t}
- \frac{(3+\sqrt{-3})}{2k}b_6e^{p_5t}
- \frac{(3-\sqrt{-3})}{2k}b_7e^{p_6t}.
\label{app117}
\end{eqnarray}
Using (\ref{app115}) and (\ref{app117}),
an explicit form for $c$ can be derived
 from the relation (\ref{app106}):
\begin{eqnarray}
c(t) & = & \frac{(-3+\sqrt{-3})}{2k}p_1b_2e^{p_1t}
- \frac{(3+\sqrt{-3})}{2k}p_2b_3e^{p_2t}
- \frac{(3+\sqrt{-3})}{2k}p_3b_4e^{p_3t}
\nonumber \\
& &
+ \frac{(-3+\sqrt{-3})}{2k}p_4b_5e^{p_4t}
+ \frac{(-3+\sqrt{-3})}{2k}p_5b_6e^{p_5t}
- \frac{(3+\sqrt{-3})}{2k}p_6b_7e^{p_6t}.
\label{app118}
\end{eqnarray}
Now with the known expressions for the functions $a,b$ and $c$ one can deduce
the form of $d(t)$ through the relation (\ref{app107}), namely,
\begin{eqnarray}
d(t) & = & -\frac{3\lambda_2}{k}b_1+
\frac{m_2}{2k_1^2}p_1^2b_2e^{p_1t}+\frac{m_1}{2k_1^2}p_2^2b_3e^{p_2t}
+\frac{m_1}{2k_1^2}p_3^2b_4e^{p_3t}
+\frac{m_2}{2k_1^2}p_4^2b_5e^{p_4t}
\nonumber\\
 & & \quad
 +\frac{m_2}{2k_1^2}p_5^2b_6e^{p_5t}+\frac{m_1}{2k_1^2}p_6^2b_7e^{p_6t},
\label{app119}
\end{eqnarray}
where
\begin{equation}
m_1=\frac{(3+\sqrt{-3})}{2k^2}, m_2=\frac{(3-\sqrt{-3})}{2k^2}.
\label{app120}
\end{equation}
For further analysis we redefine the constants
\begin{eqnarray}
r_1=\frac{m_2}{2k_1^2}p_1^2, \quad r_2=\frac{m_1}{2k_1^2}p_2^2, \quad r_3=\frac{m_1}{2k_1^2}p_3^2 \nonumber\\
r_4=\frac{m_2}{2k_1^2}p_4^2,  \quad r_5=\frac{m_2}{2k_1^2}p_5^2, \quad r_6=\frac{m_1}{2k_1^2}p_6^2.
\label{app121}
\end{eqnarray}
and write the function $d(t)$ in the form
\begin{equation}
d(t)=-\frac{3\lambda_2}{k}b_1+r_1b_2e^{p_1t}+r_2b_3e^{p_2t}+r_3b_4e^{p_3t}
+r_4b_5e^{p_4t}+r_5b_6e^{p_5t}+r_6b_7e^{p_6t}.
\end{equation}
\section{Method of solving the determining Eq.~(\ref{addcon10})}
\setcounter{section}{2}
Here we solve the  equations
\begin{eqnarray}
&&3\ddot{b}+k\dot{a}+3\lambda_1 b+kc   =  0,
\label{app201}\\
&&2\dot{c}-\ddot{a}+3\lambda_2 b +kd  =  0,
\label{app202}\\
&&\ddot{c}+2\lambda_1\dot{a}+k\dot{d} =  0,
\label{app203}\\
&&\ddot{d}-\lambda_2c+2\lambda_2\dot{a}+\lambda_1d  =  0
\label{app204}\\
&&\dddot{b}+k\lambda_2b+\lambda_1 \dot{b}+
k\dot{c}+\frac{k^2}{3}d  =  0,
\label{app205}
\end{eqnarray}
and deduce the general form of $a(t),\;b(t),\;c(t)$ and $d(t)$.

From (\ref{app201}) and (\ref{app202}) we get
\begin{eqnarray}
c & = & - \frac{3}{k}\ddot{b} - \frac{3}{k}\lambda_1b-\dot{a},
\label{app206}\\
d & = & \frac{6}{k^2}\dddot{b}+\frac{6}{k^2}\lambda_1\dot{b}
+\frac{3}{k}\ddot{a}-\frac{3\lambda_2}{k}b.
\label{app207}
\end{eqnarray}
By substituting (\ref{app206}) and (\ref{app207}) and
their derivatives into (\ref{app203}) and (\ref{app204})
 we obtain
\begin{eqnarray}
&& 3\frac{d^4b}{dt^4}+3\lambda_1\frac{d^2b}{dt^2}-3k\lambda_2\frac{db}{dt}+
2k\frac{d^3a}{dt^3}+2k\lambda_1\frac{da}{dt} = 0,
\label{app208}\\
&&2\frac{d^5b}{dt^5}+4\lambda_1\frac{d^3b}{dt^3}+2\lambda_1^2\frac{db}{dt}
+k\frac{d^4a}{dt^4}+k\lambda_1\frac{d^2a}{dt^2}+
\lambda_2k^2\frac{da}{dt} = 0.
\label{app209}
\end{eqnarray}
From (\ref{app208}) we have
\begin{eqnarray}
\frac{d^3a}{dt^3}+\lambda_1\frac{da}{dt} = -\frac{3}{2k_1}\frac{d^4b}{dt^4}
- \frac{3\lambda_1}{2k}\frac{d^2b}{dt^2}
+\frac{3\lambda_2}{2}\frac{db}{dt}.
\label{app210}
\end{eqnarray}
Substitute (\ref{app210}) into (\ref{app209}) and rearranging
we get
\begin{eqnarray}
\frac{da}{dt} = -\frac{1}{2k^2\lambda_2}\frac{d^5b}{dt^5}
-\frac{5\lambda_1}{2k^2\lambda_2}\frac{d^3b}{dt^3}
-\frac{3}{2k}\frac{d^2b}{dt^2}
-\frac{2\lambda_1^2}{k^2\lambda_2}\frac{db}{dt}.
\label{app212}
\end{eqnarray}
Integration of  (\ref{app212}) leads to
\begin{eqnarray}
a(t) = a_1-\frac{1}{2k^2\lambda_2}\frac{d^4b}{dt^4}
-\frac{5\lambda_1}{2k^2\lambda_2}\frac{d^2b}{dt^2}
-\frac{3}{2k}\frac{db}{dt}
-\frac{2\lambda_1^2}{k^2\lambda_2}b,
\label{app213}
\end{eqnarray}
where $a_1$ is an integration constant.  Substituting (\ref{app212})
and its derivatives into (\ref{app208}) we arrive at
\begin{eqnarray}
\frac{d^7b}{dt^7}+6\lambda_1\frac{d^5b}{dt^5}+9\lambda_1^2\frac{d^3b}{dt^3}
+(4\lambda_1^3+3k^2\lambda_2^2)\frac{db}{dt} = 0.
\label{app214}
\end{eqnarray}
Introducing $\frac{db}{dt} = P$, where $P$ is the new dependent variable,
we can reduce an order of the Eq.~(\ref{app214}), that is,
\begin{eqnarray}
\frac{d^6P}{dt^6}+6\lambda_1\frac{d^4P}{dt^4}+9\lambda_1^2\frac{d^2P}{dt^2}
+(4\lambda_1^3+3k^2\lambda_2^2)P = 0.
\label{app215}
\end{eqnarray}
Eq.~(\ref{app215}) is a sixth order linear ODE with constant coefficients
whose solution can be found in the following way. Since Eq.~(\ref{app215})
contains constant coefficients a general solution of this equation can be
written of the form
\begin{equation}
P = P_1e^{m_1t}+P_2e^{m_2t}+P_3e^{m_3t}+P_4e^{m_4t}+P_5e^{m_5t}
+P_6e^{m_6t},
\label{soll1}
\end{equation}
where $P_i$,$\;i=1,\ldots,6$, are integration constants and $
m_i's$,$\;i=1,\ldots,6$, are roots of the characteristic equation
\begin{eqnarray}
m^6+6\lambda_1m^4+9\lambda_1^2m^2+(4\lambda_1^3+3k^2\lambda_2^2) = 0.
\label{app216}
\end{eqnarray}
Now choosing $m^2=X$ we can rewrite the sixth power polynomial equation
(\ref{app216}) as a cubic polynomial equation in $X$, namely,
\begin{eqnarray}
X^3+6\lambda_1X^2+9\lambda_1^2X+(4\lambda_1^3+3k^2\lambda_2^2) = 0.
\label{app217}
\end{eqnarray}
Let us choose
\begin{eqnarray}
a=6\lambda_1,\quad b=9\lambda_1^2,\quad c = 4\lambda_1^3+3k^2\lambda_2^2,
\label{app218}
\end{eqnarray}
and introduce a transformation\cite{Korn:1968}
\begin{eqnarray}
X=Y-\frac{a}{3}
\label{app219}
\end{eqnarray}
in (\ref{app217}) so that it becomes
\begin{eqnarray}
Y^3+pY+q = 0,
\label{app220}
\end{eqnarray}
where the new constants, $p$ and $q$ are related to the old constants, $a,b$
and $c$, by the following relations
\begin{eqnarray}
p=b-\frac{a^2}{3}, \quad q = \frac{2}{27}a^3-\frac{ab}{3}+c.
\label{app221}
\end{eqnarray}
The cubic equation (\ref{app220}) has one real root and two complex
conjugate roots  \cite{Korn:1968}, namely,
\begin{eqnarray}
Y_1& =& A+B, \;\; Y_{2,3} = -\frac{(A+B)}{2}\pm i\frac{\sqrt{3}(A-B)}{2},
\label{app222}
\end{eqnarray}
where
\begin{eqnarray}
A = \left(-\frac{q}{2}+\sqrt{Q}\right)^{\frac{1}{3}}, \quad
B = \left(-\frac{q}{2}-\sqrt{Q}\right)^{\frac{1}{3}}, \quad
Q = \left(\frac{p}{3}\right)^3+\left(\frac{q}{2}\right)^2,
\label{app223}
\end{eqnarray}
so that the roots of Eq.~(\ref{app217}) can now be expressed through the
relation (\ref{app219}), that is,
\begin{eqnarray}
X_1 &=& A+B-\frac{a}{3}, \quad X_2 = -\frac{(A+B)}{2}-\frac{a}{3}
+i\frac{\sqrt{3}(A-B)}{2}, \nonumber\\
X_3 &=& -\frac{(A+B)}{2}-\frac{a}{3}-i\frac{\sqrt{3}(A-B)}{2}.
\label{app224}
\end{eqnarray}
From the identity $m^2=X$, we find the roots of
the characteristic equation (\ref{app216}):
\begin{eqnarray}
 \quad \quad
& & m_{1,2} = \pm \sqrt{X_1} = \pm \sqrt{A+B-\frac{a}{3}}, \quad
m_{3,4} = \pm \sqrt{X_2} = \pm \sqrt{C+iD}, \nonumber\\
& & m_{5,6} = \pm \sqrt{X_3} = \pm \sqrt{C-iD},
\label{app225}
\end{eqnarray}
where again for simplicity we have introduced the constants $C$ and $D$ which
can be fixed from the relation
\begin{eqnarray}
C = -\frac{(A+B)}{2}-\frac{a}{3}, \quad
D = \frac{\sqrt{3}(A-B)}{2}.
\label{app226}
\end{eqnarray}
Eq.  (\ref{app225}) can be put in the following convenient way
\begin{eqnarray}
m_1 & = & +\sqrt{A+B-\frac{a}{3}},\quad
m_2  =  -\sqrt{A+B-\frac{a}{3}},
\nonumber\\
m_3 & = & (C^2+D^2)^\frac{1}{4}\left[\cos\phi+i\sin\phi\right],
\quad
m_4  =  -(C^2+D^2)^\frac{1}{4}\left[\cos\phi+i\sin\phi\right],
\nonumber\\
m_5 & = & (C^2+D^2)^\frac{1}{4}\left[\cos\phi-i\sin\phi\right],
\quad
m_6  =  -(C^2+D^2)^\frac{1}{4}\left[\cos\phi-i\sin\phi\right],
\nonumber
\end{eqnarray}
where
\begin{eqnarray}
\phi = \left(\frac{\tan^{-1}(D/C)}{2}\right).
\label{app227}
\end{eqnarray}
With the help of these $m_i's,\; i=1,\dots,6$,
a general solution for (\ref{app215}) can be written in the form
\begin{eqnarray}
P = b_2e^{\alpha_1 t}+b_3e^{-\alpha_1 t}+e^{\alpha_2 t}(b_4\cos\alpha_3t
+b_5\sin\alpha_3t)
+e^{-\alpha_2t}(b_6\cos\alpha_3t-b_7\sin\alpha_3t),
\label{app228}
\end{eqnarray}
where $b_i's,\; i=2,\ldots,7$, are integration constants and $\alpha_2,
\alpha_3$ are real and imaginary parts of $m_3,m_4,m_5$ and $m_6$ respectively,
where $\alpha_2 = \sqrt{C^2+D^2}\cos\phi,$ $ \alpha_3 = \sqrt{C^2+D^2}\sin\phi$.

From the identity $\frac{db}{dt} = P$,  the function $b(t)$
upon integration can be deduced as
\begin{eqnarray}
b(t) & = & b_1+\frac{b_2}{\alpha_1}e^{\alpha_1 t}
-\frac{b_3}{\alpha_1}e^{-\alpha_1t}
+ \frac{b_4e^{\alpha_2t}}{(\alpha_2^2+\alpha_3^2)}(\alpha_2\cos\alpha_3t
+\alpha_3\sin\alpha_3 t)\nonumber\\
& & +\frac{b_5e^{\alpha_2t}}{(\alpha_2^2+\alpha_3^2)}(\alpha_2\sin\alpha_3t
-\alpha_3\cos\alpha_3 t)
+ \frac{b_6e^{-\alpha_2t}}{(\alpha_2^2+\alpha_3^2)}(\alpha_3\sin\alpha_3 t
\nonumber\\
& &
-\alpha_2\cos\alpha_3t) + \frac{b_7e^{-\alpha_2t}}{(\alpha_2^2+\alpha_3^2)}(\alpha_2\sin\alpha_3t
+\alpha_3\cos\alpha_3 t),
\label{app229}
\end{eqnarray}
where $b_1$ is an integration constant.  Once $b(t)$ is known the function
$a(t)$ can be fixed from the relation (\ref{app213}),
\begin{eqnarray}
a(t) & = & a_1-\frac{2\lambda_1^2}{k^2\lambda_2}b_1+\beta_1b_2e^{\alpha_1 t}
+\beta_2b_3e^{-\alpha_1t}
+ b_4e^{\alpha_2t}(\beta_3\cos\alpha_3t
+\beta_4\sin\alpha_3 t)\nonumber\\
& &
+ b_5e^{\alpha_2t}(\beta_3\sin\alpha_3 t-\beta_4\cos\alpha_3t)
+b_6e^{-\alpha_2t}(\beta_5\cos\alpha_3t
+\beta_4\sin\alpha_3 t)\nonumber\\
& & +b_7e^{-\alpha_2t}(\beta_4\cos\alpha_3t
-\beta_5\sin\alpha_3 t),
\label{app230}
\end{eqnarray}
where $a_1$ is an integration constant.
For simplicity in the expression $a(t)$ we have introduced the constants
$\beta_i,\;i=1,\ldots,5$.  These new constants are related to the old
parameters through the following relations
\begin{eqnarray}
\beta_1 & = & -\left(\frac{\alpha_1^3}{2k^2\lambda_2}
          +\frac{5\lambda_1\alpha_1}{2k^2\lambda_2}
      +\frac{3}{2k}
      +\frac{2\lambda_1^2}{k^2\lambda_2\alpha_1}\right), \;\;
\beta_2 =  \left(\frac{\alpha_1^3}{2k^2\lambda_2}
          +\frac{5\lambda_1\alpha_1}{2k^2\lambda_2}
      -\frac{3}{2k}
      +\frac{2\lambda_1^2}{k^2\lambda_2\alpha_1}\right),
      \nonumber\\
\beta_3 & = & -\left(\frac{\alpha_2(\alpha_2^2-3\alpha_3^2)}{2k^2\lambda_2}
          +\frac{5\lambda_1\alpha_2}{2k^2\lambda_2}
      +\frac{3}{2k}
      +\frac{2\lambda_1^2\alpha_2}{k^2\lambda_2(\alpha_2^2+
      \alpha_3^2)}\right),
      \nonumber\\
\beta_4 & = & -\left(\frac{\alpha_3(\alpha_3^2-3\alpha_2^2)}{2k^2\lambda_2}
          -\frac{5\lambda_1\alpha_3}{2k^2\lambda_2}
      +\frac{2\lambda_1^2\alpha_3}{k^2\lambda_2(\alpha_2^2+
      \alpha_3^2)}\right),
      \nonumber\\
\beta_5 & = & \left(\frac{\alpha_2(\alpha_2^2-3\alpha_3^2)}{2k^2\lambda_2}
          +\frac{5\lambda_1\alpha_2}{2k^2\lambda_2}
      -\frac{3}{2k}
      +\frac{2\lambda_1^2\alpha_2}{k^2\lambda_2(\alpha_2^2+
      \alpha_3^2)}\right).
      \label{app231}
\end{eqnarray}
With the forms of $a(t)$ and $b(t)$ the function $c(t)$ can be deduced using
the relation (\ref{app206}) in the form,
\begin{eqnarray}
c(t) & = & -\frac{3\lambda_1}{k}b_1+\gamma_1b_2e^{\alpha_1 t}
+\gamma_2b_3e^{-\alpha_1t}
+ b_4e^{\alpha_2t}(\gamma_3\cos\alpha_3t
+\gamma_4\sin\alpha_3 t)\nonumber\\
& & + b_5e^{\alpha_2t}(\gamma_3\sin\alpha_3t-\gamma_4\cos\alpha_3t)
+ b_6e^{-\alpha_2t}(\gamma_5\cos\alpha_3t
+\gamma_6\sin\alpha_3 t)\nonumber\\
& & +b_7e^{-\alpha_2t}(\gamma_6\cos\alpha_3t
-\gamma_{5}\sin\alpha_3 t),
\label{app232}
\end{eqnarray}
where again we have defined new constants $\gamma_i's,\; i=1,\ldots,5$,
 for simplicity sake. Here too one  can fix their forms using
the following relations,
\begin{eqnarray}
\gamma_1  & =  & -\left(\frac{3\alpha_1}{k}+\frac{3\lambda_1}{k\alpha_1}
+\alpha_1\beta_1\right),\;\;\;
\gamma_2 = \left(\frac{3\alpha_1}{k}+\frac{3\lambda_1}{k\alpha_1}
+\alpha_1\beta_2\right),\label{app233} \\
\gamma_3 & = & -\left(\frac{3\alpha_2}{k}+\frac{3\lambda_1\alpha_2}
{k(\alpha_2^2+\alpha_3^2)}+\alpha_2\beta_3+\alpha_3\beta_4\right),\;\;
\gamma_4  =  \left(\frac{3\alpha_3}{k}-\frac{3\lambda_1\alpha_3}
{k(\alpha_2^2+\alpha_3^2)}+\alpha_3\beta_3-\alpha_2\beta_4\right), \nonumber\\
\gamma_5  & = & \left(\frac{3\alpha_2}{k}+\frac{3\lambda_1\alpha_2}
{k(\alpha_2^2+\alpha_3^2)}+\alpha_2\beta_5-\alpha_3\beta_4\right),\;\;
\gamma_6 =  \left(\frac{3\alpha_3}{k}-\frac{3\lambda_1\alpha_3}
{k(\alpha_2^2+\alpha_3^2)}+\alpha_2\beta_4+\alpha_3\beta_5\right).
\nonumber
\end{eqnarray}
Finally, the function $d(t)$ can be derived from the relation (\ref{app207})
by simply substituting the forms  of $a(t)$ and $b(t)$ and their
derivatives into it, that is,
\begin{eqnarray}
d(t) & = & -\frac{3\lambda_2}{k}b_1+\delta_1b_2e^{\alpha_1 t}
+\delta_2b_3e^{-\alpha_1t}
+ b_4e^{\alpha_2t}(\delta_3\cos\alpha_3t
+\delta_4\sin\alpha_3 t)\nonumber\\
& &
+ b_5e^{\alpha_2t}(\delta_3\sin\alpha_3t-\delta_4\cos\alpha_3t)
+ b_6e^{-\alpha_2t}(\delta_5\cos\alpha_3t
+\delta_6\sin\alpha_3 t)\nonumber\\
&&
+b_7e^{-\alpha_2t}(\delta_6\cos\alpha_3t
-\delta_{5}\sin\alpha_3 t).
\label{app234}
\end{eqnarray}
Again for simplicity we have defined the constants $\delta_i's,\;i=1,\ldots,
6$, which can be determined from the following relations,
\begin{eqnarray}
\delta_1 & = & \frac{1}{k}\left(\frac{6\alpha_1^2}{k}+\frac{6\lambda_1}{k}
+3\alpha_1^2\beta_1-\frac{3\lambda_2}{\alpha_1}\right),\;\;
\delta_2  =  \frac{1}{k}\left(\frac{6\alpha_1^2}{k}+\frac{6\lambda_1}{k}
+3\alpha_1^2\beta_1+\frac{3\lambda_2}{k}\right),
\nonumber\\
\delta_3 & = & \frac{1}{k}\left(3(\alpha_2^2-\alpha_3^2)(\beta_3+\frac{2}{k})
+\frac{6\lambda_1}{k}+6\alpha_2\alpha_3\beta_4-\frac{3\alpha_2\lambda_2}
{(\alpha_2^2+\alpha_3^2)}\right),
\nonumber\\
\delta_4 & = & -\frac{1}{k}\left(6\alpha_2\alpha_3(\beta_3+\frac{2}{k})
-3\beta_4(\alpha_2^2-\alpha_3^2)
+\frac{3\alpha_3\lambda_2}{(\alpha_2^2+\alpha_3^2)}\right),
\nonumber\\
\delta_5 & = & \frac{1}{k}\left(3(\alpha_2^2-\alpha_3^2)(\beta_5+\frac{6}{k})
 +\frac{6\lambda_1}{k}-6\alpha_2\alpha_3\beta_4+\frac{3\alpha_2\lambda_2}
{(\alpha_2^2+\alpha_3^2)}\right),
\nonumber\\
 \quad \quad
\delta_6 & = & \frac{1}{k}\left(6\alpha_2\alpha_3(\beta_5+\frac{2}{k})
+3\beta_4(\alpha_2^2-\alpha_3^2)-
\frac{3\lambda_2\alpha_3}{(\alpha_2^2+\alpha_3^2)}\right).
\label{app235}
\end{eqnarray}

%\end{appendix}

\end{document}